\documentclass[useAMS,usenatbib,a4paper]{mn2e}
\usepackage{amsmath}
\usepackage[pdftex]{graphicx}
\usepackage{epstopdf}
\usepackage[english]{babel}
\usepackage{subfigure}
\usepackage{array,multirow}
\usepackage{underscore}
\usepackage{natbib}
\usepackage{float}
\usepackage{amssymb}
\usepackage{dcolumn}

\newcolumntype{L}[1]{>{\raggedright\let\newline\\\arraybackslash\hspace{0pt}}m{#1}}
\newcolumntype{C}[1]{>{\centering\let\newline\\\arraybackslash\hspace{0pt}}m{#1}}
\newcolumntype{R}[1]{>{\raggedleft\let\newline\\\arraybackslash\hspace{0pt}}m{#1}}

\def\mnras{MNRAS}
\def\apj{ApJ}
\def\aj{AJ}

\def\apjl{ApJL}
\def\apjs{ApJS}

\def\pasp{PASP}
\def\nat{Nature}
\def\pasa{PASA}

\title[Red nuggets growing inside-out]{Evidence for the inside-out growth of red nuggets from gravitational lensing}
\title[Red nuggets growing inside-out]{Red nuggets, growing inside-out}
\title[Red nuggets growing inside-out]{Red nuggets grow inside-out: evidence from gravitational lensing}

\author[L. J. Oldham et al.]{Lindsay Oldham$^{1}$\thanks{E-mail: loldham@ast.cam.ac.uk}, Matthew W.~Auger$^{1}$, Christopher D. Fassnacht$^{2}$, Tommaso Treu$^{3}$,
\and Brendon J.~Brewer$^{4}$, L.V.E. Koopmans$^{5}$, David Lagattuta$^{6}$, Philip Marshall$^{7}$, 
\and John McKean$^{8,9}$, Simona Vegetti$^{10}$ \\
$^{1}$ Institute of Astronomy, University of Cambridge, Madingley Road, Cambridge CB3 0HA, UK \\
$^{2}$ Department of Physics, University of California, Davis, 1 Shields Ave. Davis, CA 95616, USA\\
$^{3}$ Department of Physics and Astronomy, UCLA, 430 Portola Plaza, Los Angeles, CA 90095-1547, USA\\
$^{4}$ Department of Statistics, The University of Auckland, Private Bag 92019, Auckland 1142, New Zealand\\
$^{5}$ Kapteyn Astronomical Institute, University of Groningen, P.O. Box 800, 9700 AV Groningen, The Netherlands\\
$^{6}$ Universit\'e de Lyon, Universit\'e de Lyon, CNRS, Centre de Recherche Astrophysique de Lyon UMR5574, F-69230, Saint-Genis-Laval, France\\
$^{7}$ Kavli Institute for Particle Astrophysics and Cosmology, Stanford University, 452 Lomita Mall, Stanford, CA 94035, USA\\
$^{8}$ ASTRON, Netherlands Institute for Radio Astronomy, Postbus 2, NL-7990 AA, Dwingeloo, the Netherlands\\
$^{9}$ Kapteyn Astronomical Institute, PO Box 800, NL-9700 AV Groningen, the Netherlands \\
$^{10}$ Max Planck Institute for Astrophyiscs, Karl-Schwarzschild-Strasse 1, D-85740 Garching, Germany\\
}

\begin{document}
\maketitle
\setcounter{page}{1}

\begin{abstract}
We present a new sample of strong gravitational lens systems where both the foreground lenses and background sources are early-type galaxies. Using imaging from HST/ACS and Keck/NIRC2, we model the surface brightness distributions and show that the sources form a distinct population of massive, compact galaxies at redshifts $0.4 \lesssim z \lesssim 0.7$, lying systematically below the size-mass relation of the global elliptical galaxy population at those redshifts. These may therefore represent relics of high-redshift red nuggets or their partly-evolved descendants. We exploit the magnifying effect of lensing to investigate the structural properties, stellar masses and stellar populations of these objects with a view to understanding their evolution. We model these objects parametrically and find that they generally require two S\'ersic components to properly describe their light profiles, with one more spheroidal component alongside a more envelope-like component, which is slightly more extended though still compact. This is consistent with the hypothesis of the inside-out growth of these objects via minor mergers. We also find that the sources can be characterised by red-to-blue colour gradients as a function of radius which are stronger at low redshift -- indicative of ongoing accretion -- but that their environments generally appear consistent with that of the general elliptical galaxy population, contrary to recent suggestions that these objects are predominantly associated with clusters. 
\end{abstract}


\section{Introduction}

The discovery that massive, quiescent galaxies at redshifts $z>2$ are extremely compact \citep{Daddi2005,Trujillo2006,vanDokkum2008,Damjanov2009,Damjanov2011} relative to their local counterparts has opened the door to important tests of our models of galaxy evolution. While the hierarchical paradigm allows for the growth of passive galaxies via dissipationless mergers at a rate which may be able to account for the evolution that is required at $z \lesssim 1.5$ (e.g. \citealp{Nipoti2012,Newman2012,Posti2014}, but see also \citealp{Sonnenfeld2014b}), this cannot explain the amount of evolution observed at higher redshifts or the tightness of galaxy scaling relations \citep{Shankar2013}. Adiabatic processes, such as expansion triggered by quasar feedback \citep{Fan2010}, may also be important, and the role of progenitor bias, as opposed to the growth of individual systems, remains unclear \citep{Newman2012,Carollo2013,Belli2014}. 

One potentially powerful way of distinguishing between these scenarios is to quantify the morphological evolution of these galaxies. Mergers and adiabatic expansion should each leave particular imprints on the structure and stellar populations of a galaxy \citep{Hopkins2009,Fan2010,Hilz2013}, and so it should be possible to set some constraints on their relative importance in individual systems at lower redshifts. The studies of \citet{Stockton2014} and \citet{Hsu2014} attempted this at redshifts $z \sim 0.5$, using adaptive optics (AO) imaging of small galaxy samples, and found a large fraction of flattened galaxies, suggestive of disky or prolate structures, and low S\'ersic indices, possibly consistent with the existence of accreted envelopes. However, discrepancies between stellar and dynamical masses in both studies (which could be indicative of high stellar velocity anisotropies resulting from their flattened morphologies) highlight the fact that their observations are really pushing the capabilities of our current observing facilities.

Strong gravitational lensing, however, allows massive galaxies in the Universe to act as natural telescopes. %When light from a galaxy passes close to an intervening massive object, it is deflected by the gravitational potential of the mass and seen as multiple magnified images on the sky, separated on scales of the Einstein radius $R_{Ein}$ of the mass, which is typically a few arcseconds.
Because lensing conserves surface brightness, a lensed background source galaxy appears not only larger, but also brighter, and this makes it possible to probe the light distributions of very small objects with high signal-to-noise data \citep[e.g.][]{Newton2011}. Furthermore, the magnification bias of strong lensing tends to favour compact sources, making it an ideal tool to study a population of intermediate-redshift massive, compact galaxies at much higher resolutions than would otherwise be possible.

In this paper, we present a new sample of thirteen early-type/early-type lens systems (EELs). These were identified as lens candidates using the SDSS spectroscopic database by searching for spectra that could be decomposed into two early-type galaxy (ETG) spectra at different redshifts, and confirmed using AO imaging in the $K'$-band as part of the Strong lensing at High Angular Resolution Programme \citep[SHARP;][]{Lagattuta}. These now form roughly half of the SHARP sample, and, in addition to the source science presented here, will also be a critical resource for SHARP's ongoing substructure investigations \citep[e.g.][]{Vegetti2012}. The first EEL has already been shown to be a massive, compact ETG at redshift $z = 0.63$, and was found to require a two-component S\'ersic model to accurately fit the surface brightness profile, including an extended low-surface-brightness component \citep{Auger2011}, in line with expectations of the effect of merging and accretion on high-redshift nuggets \citep{Hopkins2009}. However, those models were based on single-band AO imaging with an uncertain PSF (whose broad wings generally affect the measurement of the low-surface brightness outskirts); we now have Hubble Space Telescope/Advanced Camera for Studies (HST/ACS) images for all of the EELs, facilitating a much more thorough study. Here, we analyse the entire sample to investigate and exploit the idea that this relatively unexplored class of gravitational lenses naturally selects compact nugget descendants. 

The paper is structured as follows: we present the data in Section 2 and our lens modelling methods and results in Sections 3 and 4. We then investigate and discuss the properties of the source galaxies in Sections 5 and 6 and finally conclude in Section 7. Throughout the paper, we use AB magnitudes and circularised radii, calculate stellar masses assuming a Chabrier stellar initial mass function (IMF), and assume a flat $\Lambda$CDM cosmology with $\Omega_m = 0.3$ and $h = 0.7$.

%(The evolution in the number density of these objects is also a matter of debate, with the early indications from searches of the Sloan Digitial Sky Survey (SDSS) that low-redshift analogues were extremely rare \citep{Trujillo2009,Stockton2010,Taylor2010} being currently overturned by searches of the deeper, higher-quality datasets that we now have access to such as the Kilo Degree Survey (KiDS) and the COSMOS field catalogues \citep{Damjanov2015a,Tortora2016}. Initial forecasts from these studies imply that the evolution of the number density of these systems evolves only mildly at low redshifts, and does not deviate significantly from the predictions of semi-analytic models and hydrodynamical simulations accounting for their evolution through minor mergers \citep[e.g.][]{Quilis2013, Wellons2016}.)

%However, at At high redshifts, their distance from us and intrinsically small sizes conspire to make it difficult to measure anything beyond their most general properties \citep[though see][]{vanDokkum2009,Newman2010}.

\section{Data}

As summarised by \citet{Auger2011}, EEL candidates were identified by searching the SDSS spectroscopic database for spectra that could be decomposed into two ETG spectra at different redshifts \citep[similarly to the method emloyed by the Sloan Lens ACS survey, SLACS;][though the SLACS survey searched for emission lines in the background sources]{Bolton2006}. SDSS imaging was used to reject lens candidates that were clearly resolved into two galaxies, and a probability for lensing was determined based upon the velocity dispersion of the foreground galaxy. Fourteen candidates were observed in the $K'$-band using NIRC2 with laser guide star adaptive optics (LGS-AO) on Keck II over a range of dates from August 2009 until May 2012, most as part of SHARP, and all were confirmed as lenses. The data were reduced as described by \citet{Auger2011}, with images taken using the wide camera drizzled to a scale 0.03$''$/pixel and those taken using the narrow camera drizzled to a scale of 0.01$''$/pixel. The zeropoints for these data were calibrated against 2MASS photometry, which includes robust detections of all of the systems except J0913 and J1446. For these two objects, we used observations of other targets observed on the same nights and determined zeropoints for these based upon 2MASS photometry, finding negligible scatter throughout the nights.

These EELs were also observed using HST/ACS as part of the programme GO 13661 (PI: Auger). Two dithered exposures of duration $\sim$500~s were observed in the $I$-band (F814W), and another set of two dithered exposures of $\sim$500~s were obtained in the $V$-band (F555W for sources at redshift $z<0.55$ or F606W for $z>0.55$, in order to straddle the 4000\AA~break). The ACS data were reduced using {\sc Astrodrizzle} and were drizzled to a scale of $0.05''$/pixel. There are a small number of artefacts in the resulting images due to the limited number of exposures in each band, and these are masked in the subsequent analysis. The positions on the sky of these fourteen systems are summarised in Table 1, along with the redshifts of both source and lens.

\begin{figure*}
\centering
\subfigure{\includegraphics[trim = 18 18 18 18,clip,width=\textwidth]{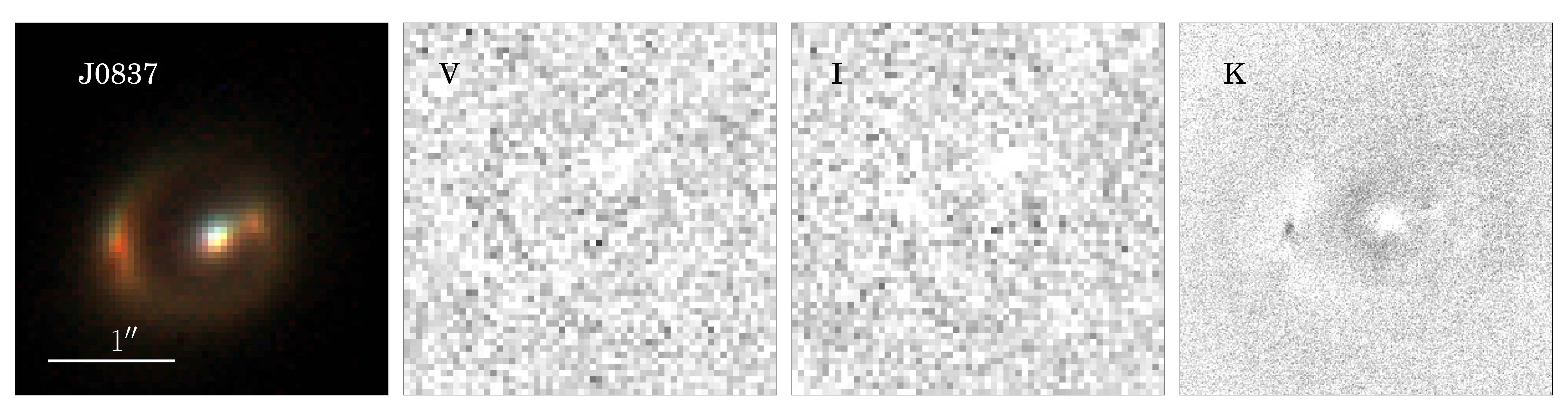}}\hfill
\subfigure{\includegraphics[trim = 18 18 18 18,clip,width=\textwidth]{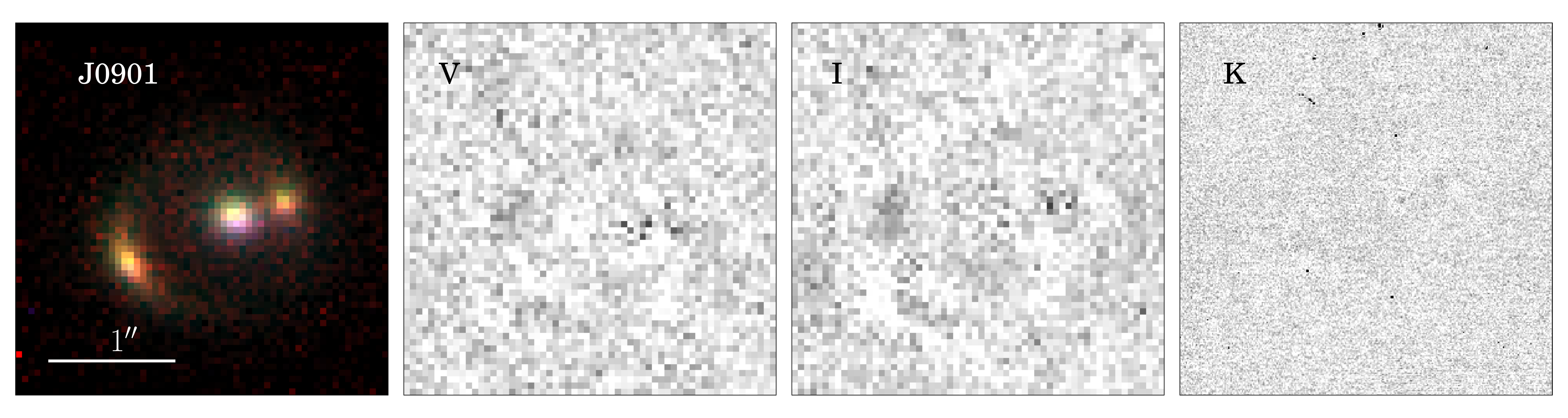}}\hfill
\subfigure{\includegraphics[trim = 18 18 18 18,clip,width=\textwidth]{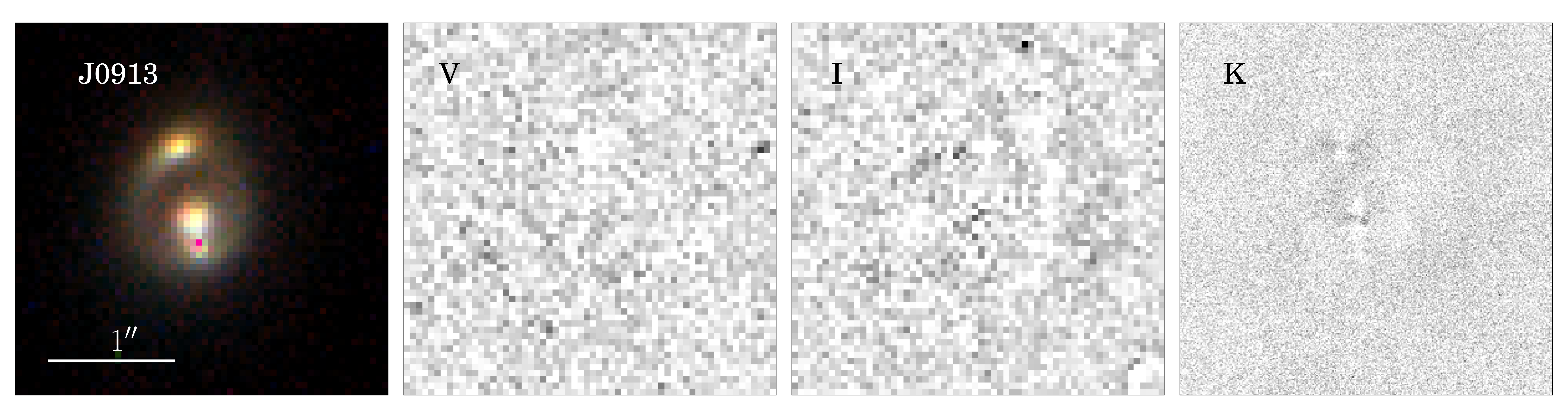}}\hfill

\subfigure{\includegraphics[trim = 18 18 18 18,clip,width=\textwidth]{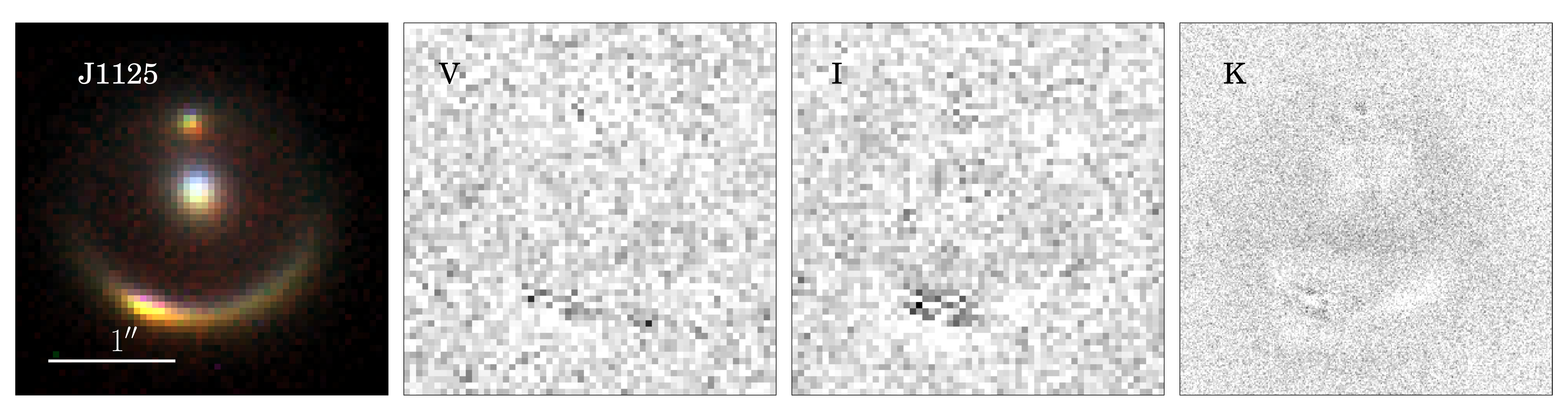}}\hfill

\caption{From left to right, we show the colour image combining all three bands of data and the residuals for the $V$, $I$ and $K'$ bands, for the best model (i.e., 1C/2C) for each system as given in Table 2. All cutouts are 3 arcseconds on a side. }
\label{fig:eels}
\end{figure*}

\begin{figure*}
\subfigure{\includegraphics[trim = 18 18 18 18,clip,width=\textwidth]{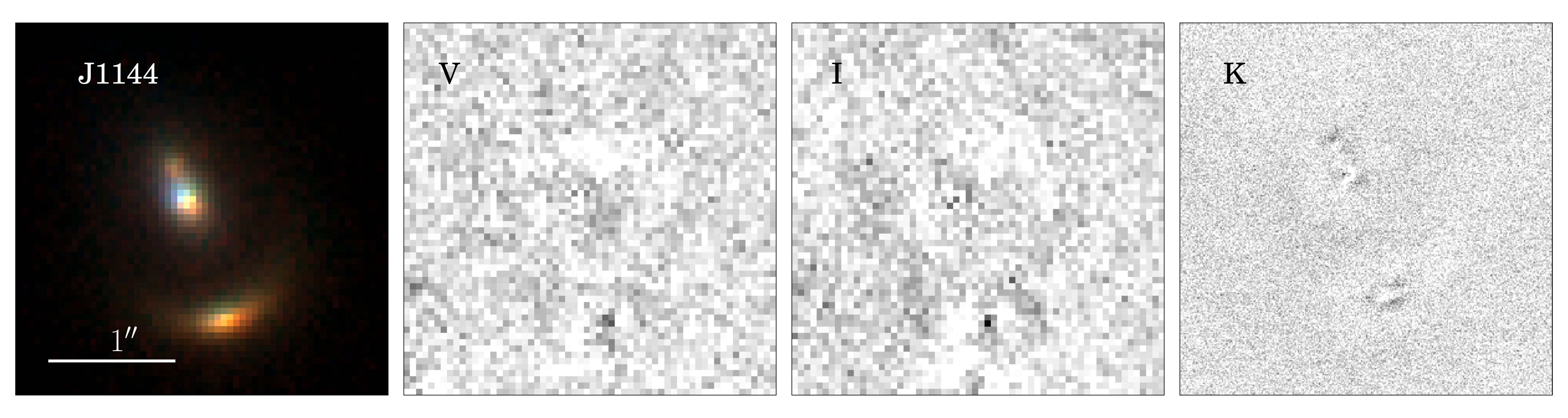}}\hfill
\subfigure{\includegraphics[trim = 18 18 18 18,clip,width=\textwidth]{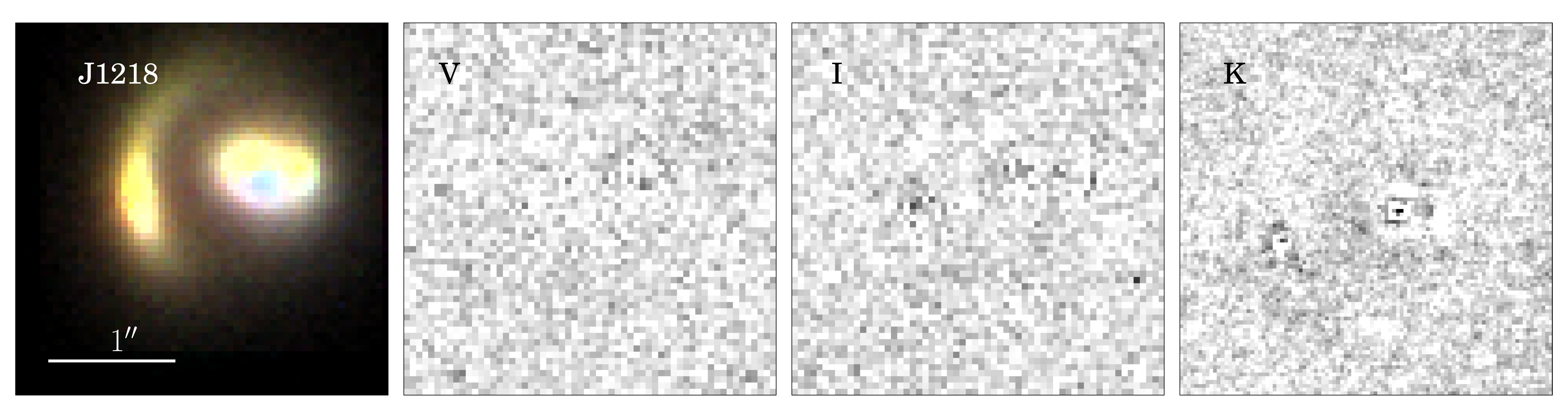}}\hfill
\subfigure{\includegraphics[trim = 18 18 18 18,clip,width=\textwidth]{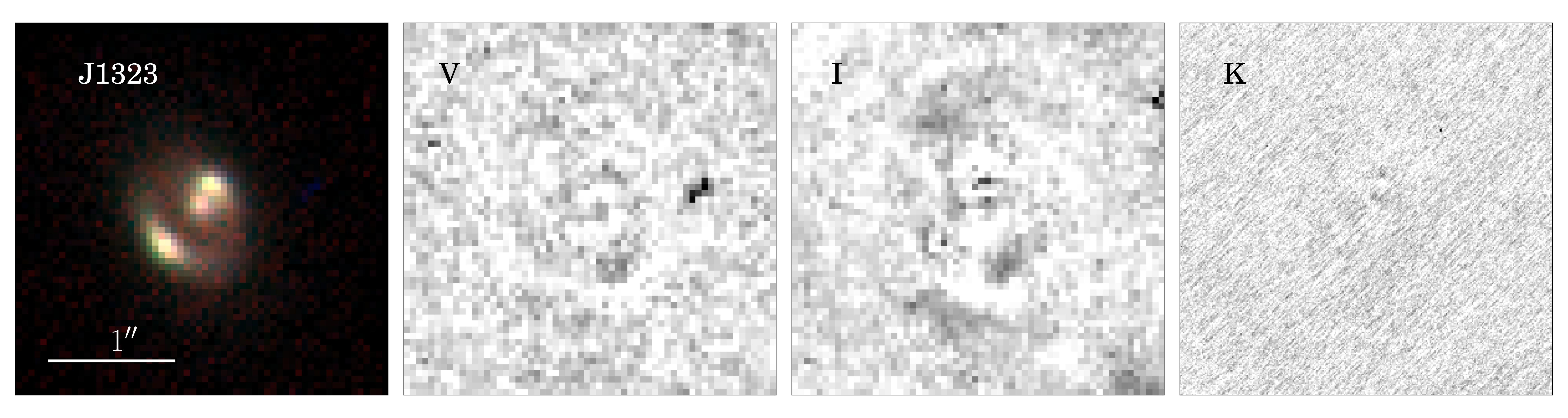}}\hfill
\subfigure{\includegraphics[trim = 18 18 18 18,clip,width=\textwidth]{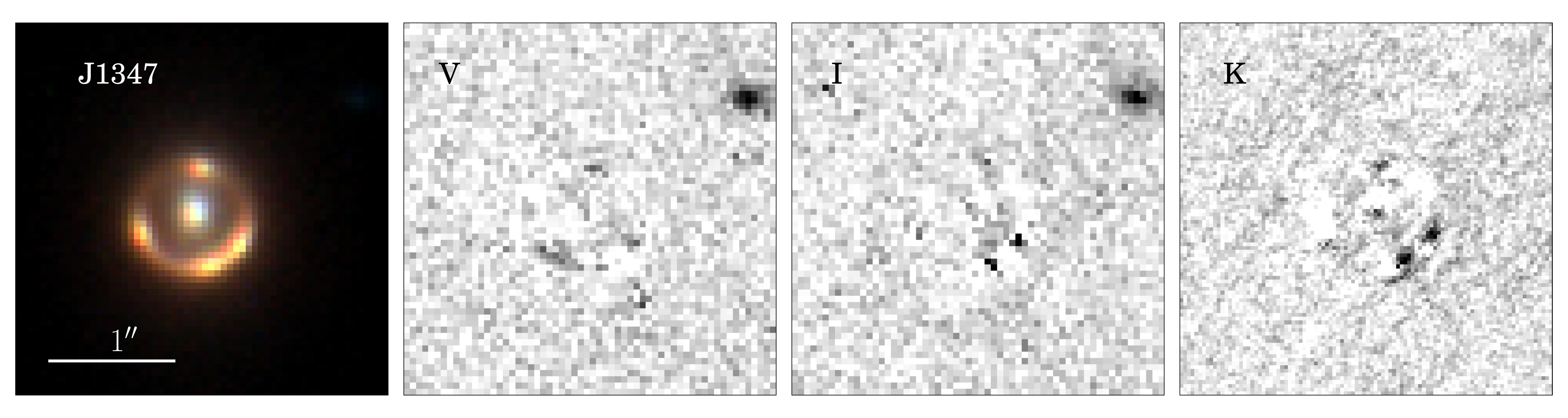}}\hfill
\subfigure{\includegraphics[trim = 18 18 18 18,clip,width=\textwidth]{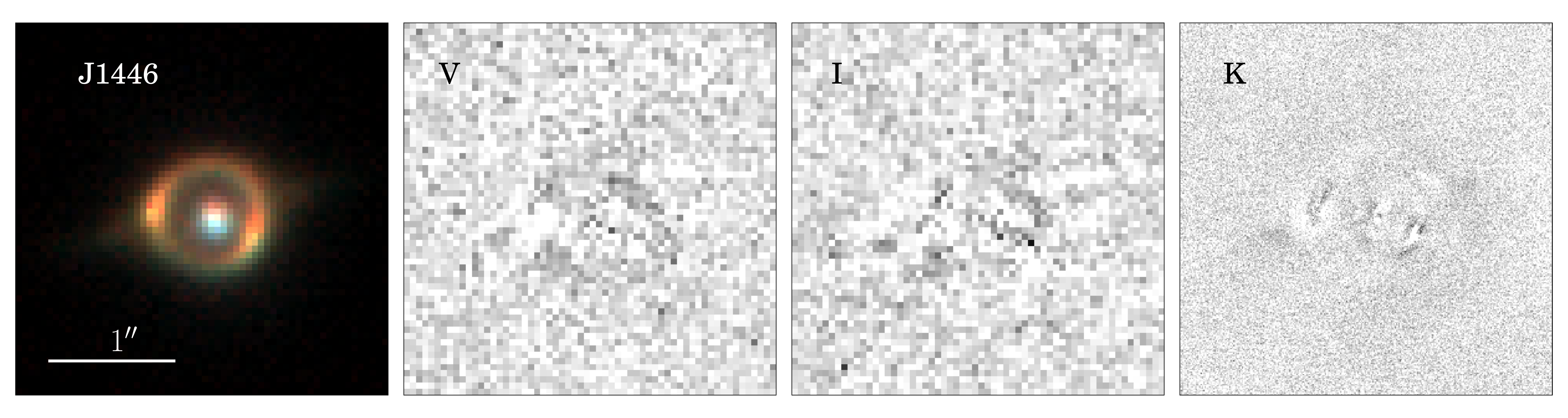}}\hfill
\contcaption{}
\end{figure*}

\begin{figure*}
\subfigure{\includegraphics[trim = 18 18 18 18,clip,width=\textwidth]{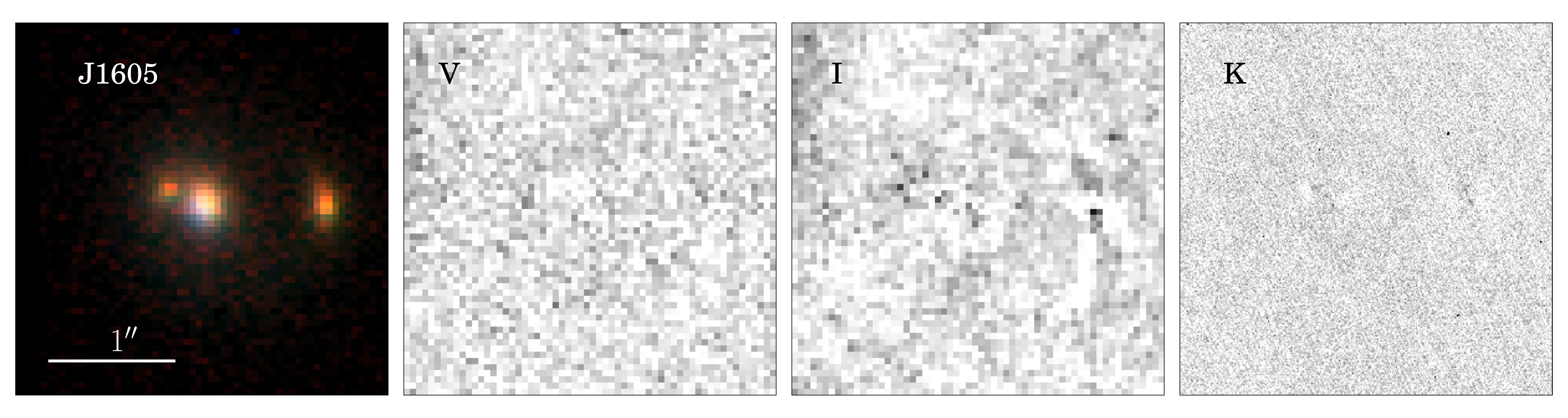}}\hfill
\subfigure{\includegraphics[trim = 18 18 18 18,clip,width=\textwidth]{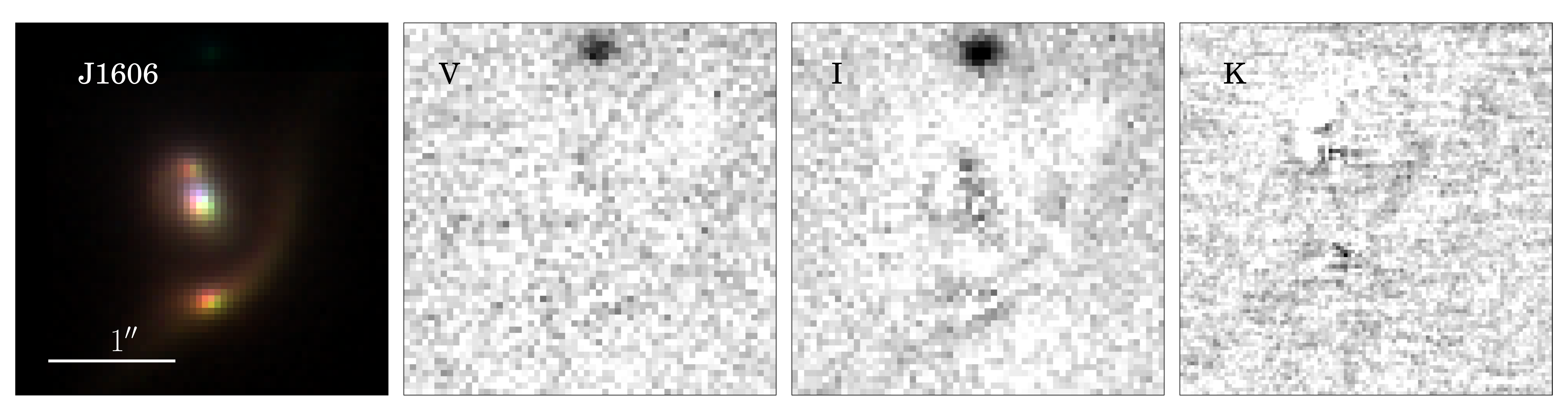}}\hfill
\subfigure{\includegraphics[trim = 18 18 18 18,clip,width=\textwidth]{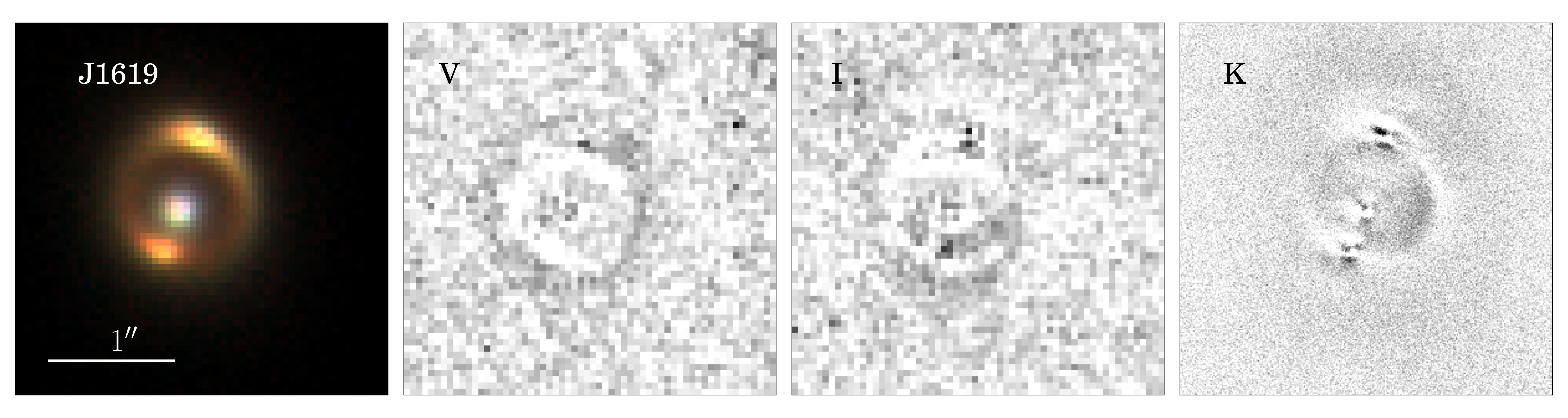}}\hfill
\subfigure{\includegraphics[trim = 18 18 18 18,clip,width=\textwidth]{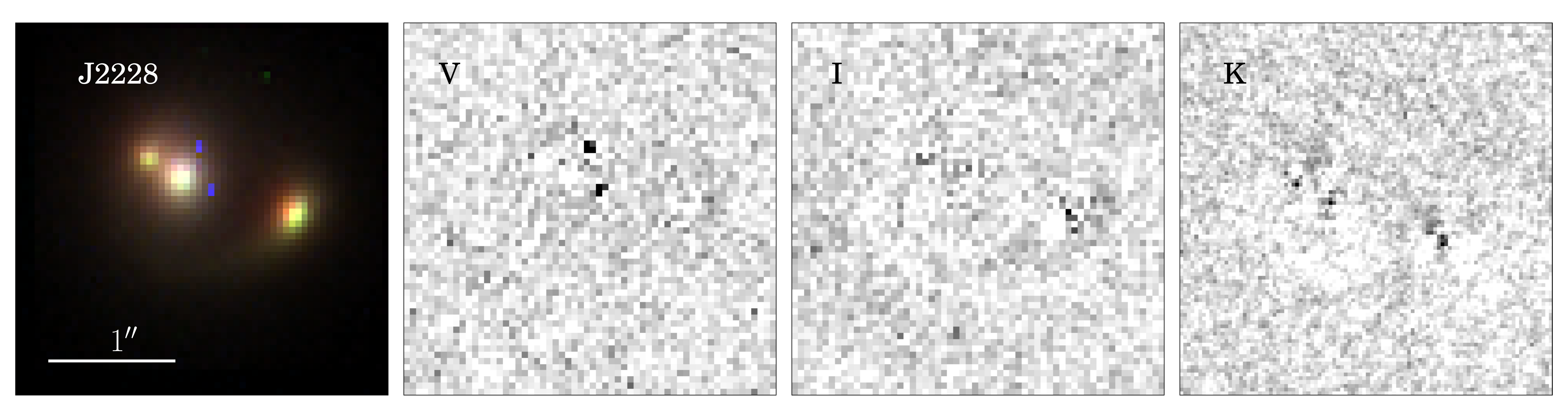}}\hfill
\contcaption{}
%\caption{EELs models ctd. From left to right, we show the colour image combining all three bands of data and the residuals for the $V$, $I$ and $K'$ bands, for the best model (i.e. 1C/2C) for each system as given in Table 2. All cutouts are 3 arcseconds on a side. }
%\label{fig:eels}
\end{figure*}

\

%%% TABLE OF RA, DEC, ZL, ZS
\begin{table}\
\centering
\begin{tabular}{ccccc}\hline
 EEL & RA (deg) & DEC (deg) & $z_l$ & $z_s$ \\\hline
J0837 &  08:37:01.21 &  +08:01:17.89   &  0.4248 &  0.6406 \\
J0901 &  09:01:21.25 &  +20:27:40.41   &  0.3108 &  0.5860 \\
J0913 &  09:13:45.65 &  +42:37:30.81   &  0.3946 &  0.5390 \\
J1125 &  11:25:13.89 &  +30:58:05.59   &  0.4419 &  0.6884 \\
J1144 &  11:44:28.40 &  +15:40:39.36   &  0.3715 &  0.7050 \\
J1218 &  12:18:06.67 &  +56:48:05.12   &  0.3177 &  0.6000 \\
J1248 &  12:48:47.82 &  +47:11:05.81   &  0.3042 &  0.5276 \\
J1323 &  13:23:59.07 &  +39:46:33.24   &  0.3192 &  0.4637 \\
J1347 &  13:47:04.96 &  $-$01:01:03.57 &  0.3974 &  0.6289 \\
J1446 &  14:46:30.20 &  +38:56:56.41   &  0.3175 &  0.5858 \\
J1605 &  16:05:23.28 &  +38:11:53.95   &  0.3065 &  0.5418 \\
J1606 &  16:06:07.09 &  +22:35:11.35   &  0.3810 &  0.6545 \\
J1619 &  16:19:12.63 &  +20:24:27.97   &  0.3635 &  0.6132 \\
J2228 &  22:28:40.80 &  $-$00:18:16.84 &  0.2387 &  0.4366 \\\hline

\end{tabular}
\caption{Positions and redshifts, for both source and lens, of the fourteen EELs.}
\end{table}

\section{Lens modelling}
\label{sec:lensmodelling}

One of the main aims of this study is to robustly measure the sizes, morphologies and masses of the source galaxies in order to compare their size-mass relation with both other galaxies at similar redshifts and high-redshift nuggets; we therefore choose to model their light distributions using elliptical S\'ersic profiles. An alternative would be to make pixellated source reconstructions \citep[e.g.][]{Warren2003,TreuKoopmans2004, Koopmans2005, Vegetti2009} from which half-light radii could be measured. However, this would add an extra level of uncertainty to the final size and magnitude measurements and complicate the interpretation of the sizes; nevertheless, for a small number of systems, we do carry out inference based on pixellated sources as a verification of our parametric lens models, but we do not use these in the analysis. (We also make pixellated reconstructions of all the EELs sources, and show these in the Appendix.) Further, single-component S\'ersic profiles are a standard way of modelling surface brightness distributions for both lensed and unlensed galaxies at all redshifts \citep[e.g.][]{Shen2003, Newton2011, vanderWel2014}, so modelling our lensed sources in a similar way allows a straightforward comparison with other studies \citep[see][for a discussion of the advantages and limitations of parametric source modelling]{Marshall2007}.

Equally, some sources with more complex light distributions may not be well described by single-component S\'ersic models -- for instance, those containing bars or bulges and disks -- and, from a lensing point of view, it is important to verify that any residuals in the model are a result of the shortcomings of the light profile that has been imposed, rather than the mass model. Further, it is important to be able to measure the total flux from the source and assess any uncertainty or bias introduced by assuming a single S\'ersic profile. For each system, we therefore create two `best' models, the first using a single S\'ersic component for the source (which we call a 1C or `one-component' model) and the second with two S\'ersic components (which we call a 2C or `two-component' model); for some systems, the 1C model allows us to describe the data down to the noise level, and we do not create 2C models in these cases. For the foreground galaxy, we also use either one or two components. In all models with more than one component for either the foreground galaxy or the source, we require the two components to be concentric, but allow their position angles and ellipticities to be independent.

For each first S\'ersic component, we therefore have six free non-linear parameters -- $(x,y,q_1,\phi_1,R_{e,1}, n_1)$ -- where ($x$, $y$) gives the centroid, $q_1, \phi_1$ describe the axis ratio and position angle and $R_{e,1}, n_1$ describe the half-light radius and index of the S\'ersic profile. For each second S\'ersic component, we have four free parameters: $(q_2,\phi_2,R_{e,2},n_2)$. We model the lensing mass of the foreground galaxy using an elliptical power law distibution \citep[calculating deflection angles according to the prescription of][]{Barkana1998} and allow for an external shear; while the simpler, more common singular isothermal ellipsoid (SIE) distribution has been shown to provide a good approximation to the lens potential on galaxy scales \citep[e.g.][]{TreuKoopmans2004}, our focus is on measuring reliable and robust sizes and we therefore want to eliminate as much potential bias in our source models as possible. Our mass model therefore has eight free parameters -- $(x_l,y_l,q_l,\phi_l,R_{Ein},\eta,\gamma_{ext}, \phi_{ext})$ -- where $(x_l, y_l)$ describe the centroid of the mass, $q_l,\phi_l$ give its axis ratio and position angle, $R_{Ein}, \eta$ give the Einstein radius and the power law index of the 3D density profile $\rho \propto r^{-(\eta+1)}$ and $\gamma_{ext},\phi_{ext}$ give the magnitude and position angle of an external shear. We do not require the mass and light of the lens galaxy to be concentric or aligned. 

For a given set of these \textit{non-linear} parameters, we determine the \textit{linear} amplitude of each surface brightness component by evaluating the foreground galaxy profile in the image plane and the source galaxy profile in the source plane, given the deflection angles of the mass model. We do not subtract the foreground galaxy light prior to the modelling due to the covariance between the foreground and background light. These are especially covariant in the EELs as compared to other lens systems due to their generally small Einstein radii and similar colours, which result in a very large amount of overlap between the source and lens light. 

The model is then convolved with the point-spread function (PSF); for the HST images, we use a nearby unsaturated star for the PSF in each band, whereas for the $K'$ band data, with an unstable PSF and often with no reference star in the field of view, we model the PSF as the sum of three (concentric but not aligned) elliptical Gaussian profiles, and infer the properties of these Gaussians along with the other model parameters. We then use a non-negative least squares linear inversion to find the best combination of the foreground lens and background source light components and a uniform background component, and thereby calculate the likelihood for the data $\vec{D}$, given the non-linear parameters of the model $\vec{M}$, as

\begin{equation}
 \ln L(\vec{D}|\vec{M}) = -\frac{1}{2}\sum_i \Big(\frac{d_i - m_i}{\sigma_i}\Big)^2
\end{equation}
where $d_i, m_i, \sigma_i$ are the i$^{th}$ pixel in the data image, model image and noise map respectively, and the sum is over all unmasked pixels (for some systems, bright interloping objects must be masked by hand). Given uniform priors on all the non-linear parameters, we can then infer the posterior distribution, $p(\vec{M}|\vec{D})$, of the model given the data in a Bayesian way using a Markov Chain Monte Carlo (MCMC) exploration. To ensure that the parameter space is fully explored when the posterior is not necessarily uni-modal, we use the parallel-tempered version of \texttt{emcee} \citep{ForemanMackey2013} with three temperatures.

We begin by modelling the HST $V$ and $I$ bands jointly, requiring the light and mass profiles to be the same in both filters (accounting for their different PSFs and spatial shifts between bands), and allowing each surface brightness component to contribute a different amount to the flux in each filter. We then model the $K'$ band separately, fixing the lensing galaxy's mass and light profiles to those inferred from the HST data and inferring the PSF and the source profile. The impetus for remodelling the source in the $K'$ band, but not the foreground galaxy, is that we are particularly interested in the structure of the potentially nugget-like source galaxies here, including the possibility that they might exhibit strong colour gradients due to ongoing or recent evolution, which would lead to smaller measured sizes in the $K'$-band. We test this rationale by creating models for a subset of the EELs in which we also fix the source profile, and infer just the PSF, and find that the residuals are considerably worse in a number of cases. We additionally create models in which the foreground galaxy light profile is also allowed to change (though the mass remains fixed), and recover a posterior distribution that is consistent with the HST models. In our analysis of the size-mass relation, we opt to use the sizes from the HST data, as these are generally more robust since they are not dependent on any inference on the PSF. 

We also create models in which the three bands are fitted simultaneously. In this case, we infer the lens mass and light profiles, which are the same in all filters, as well as the $K'$ band PSF and the source profile, where the latter is now a single S\'ersic component with a wavelength-dependent effective radius given by

\begin{equation}
 \log (R_e/\textrm{arcsec}) = \alpha_{R} \log (\lambda/6000\textrm{\AA}) + \beta_{R}
\end{equation}
for wavelength $\lambda$. This model therefore allows for colour gradients while modelling all three bands in a consistent way, and provides an important consistency check for our inferred mass profiles. It is also informative as a further way of distinguishing between different red nugget growth scenarios \citep[e.g.][]{Fan2010, Wuyts2010, Hilz2013, Ishibashi2013} which make distinct predictions for the extent and colours of the stellar populations that should be observed. These models are treated separately in Section 6.3.

\begin{table*}
 \centering
\begin{tabular}{c|ccccc D{,}{\,\pm\,}{4} D{,}{\,\pm\,}{3} D{,}{\,\pm\,}{3} c}\hline
EEL & $z_l$ & $z_s$ & $R_{Ein}$ (arcsec) & $\eta$ & $q_{lens}$ & \multicolumn{1}{r}{$\phi_{lens}$ (deg)\phantom{1}} & \multicolumn{1}{r}{$\gamma_{ext}$\phantom{100}} & \multicolumn{1}{r}{$\phi_{ext}$ (deg)\phantom{1}} & N\\\hline
J0837 & 0.4248 & 0.6406 & $ 0.56 \pm 0.01 $ & $ 1.20 \pm 0.01 $ & $ 0.76 \pm 0.01 $ &    29.80 , \phantom{1}1.18  &   0.06 , 0.01  &  -116.96 , 0.59  & 1C \\
J0901 & 0.3108 & 0.5860 & $ 0.67 \pm 0.01 $ & $ 1.07 \pm 0.01 $ & $ 0.82 \pm 0.01 $ &     5.51 , \phantom{1}1.29  &  -0.04 , 0.01  &    12.35 , 1.88  & 1C \\
J0913 & 0.3946 & 0.5390 & $ 0.42 \pm 0.01 $ & $ 1.24 \pm 0.02 $ & $ 0.79 \pm 0.02 $ &  -121.25 , \phantom{1}2.24  &   0.04 , 0.01  &   -35.66 , 3.45  & 2C \\
J1125 & 0.4419 & 0.6884 & $ 0.86 \pm 0.01 $ & $ 0.96 \pm 0.02 $ & $ 0.92 \pm 0.01 $ &   112.54 , \phantom{1}1.94  &   0.08 , 0.01  &    97.52 , 0.52  & 2C \\
J1144 & 0.3715 & 0.7050 & $ 0.68 \pm 0.01 $ & $ 1.08 \pm 0.02 $ & $ 0.75 \pm 0.02 $ &   -57.30 , \phantom{1}0.90  &  -0.04 , 0.01  &    30.15 , 3.16  & 2C \\
J1218 & 0.3177 & 0.6000 & $ 0.68 \pm 0.01 $ & $ 1.11 \pm 0.01 $ & $ 0.81 \pm 0.01 $ &   -37.48 , \phantom{1}2.06  &  -0.02 , 0.01  &   -87.30 , 4.68  & 1C \\
J1323 & 0.3192 & 0.4637 & $ 0.31 \pm 0.01 $ & $ 1.01 \pm 0.01 $ & $ 0.76 \pm 0.01 $ &   -66.18 , \phantom{1}2.14  &   0.02 , 0.01  &    11.25 , 8.06  & 1C \\
J1347 & 0.3974 & 0.6289 & $ 0.43 \pm 0.01 $ & $ 1.23 \pm 0.00 $ & $ 0.62 \pm 0.00 $ &    90.29 , \phantom{1}0.38  &  -0.01 , 0.01  &   -64.55 , 4.40  & 2C \\
J1446 & 0.3175 & 0.5858 & $ 0.41 \pm 0.01 $ & $ 1.44 \pm 0.02 $ & $ 0.79 \pm 0.01 $ &   -73.50 , \phantom{1}1.31  &   0.01 , 0.01  &    66.29 , 7.27  & 2C \\
J1605 & 0.3065 & 0.5418 & $ 0.64 \pm 0.01 $ & $ 1.25 \pm 0.02 $ & $ 0.67 \pm 0.02 $ &    98.76 , \phantom{1}1.66  &   0.06 , 0.01  &   -26.92 , 3.69  & 2C \\
J1606 & 0.3810 & 0.6545 & $ 0.52 \pm 0.01 $ & $ 1.21 \pm 0.01 $ & $ 0.62 \pm 0.01 $ &   -53.91 , \phantom{1}0.96  &   0.09 , 0.01  &    25.81 , 1.73  & 2C \\
J1619 & 0.3635 & 0.6132 & $ 0.50 \pm 0.01 $ & $ 1.28 \pm 0.03 $ & $ 0.97 \pm 0.02 $ &   -68.87 ,           16.58  &  -0.06 , 0.01  &   -34.39 , 2.21  & 2C \\
J2228 & 0.2387 & 0.4366 & $ 0.60 \pm 0.01 $ & $ 1.12 \pm 0.01 $ & $ 0.96 \pm 0.01 $ &   -59.87 , \phantom{1}7.59  &  -0.06 , 0.01  &     1.59 , 2.01  & 2C \\\hline

\end{tabular}
\caption{A summary of the lens models, inferred using the HST $V$- and $I$-band data, with statistical uncertainties. We present the lens and source redshifts (measured from the SDSS spectra), the Einstein radius in arcsec, the power-law index $\eta$, the ellipticity and position angle of the lens and the magnitude and position angle of the extrnal shear. The final column denotes the `best' model for each system, which is either 1C (one S\'ersic component) or 2C (two S\'ersic components).}
\label{tab:lensmodelsY}
\end{table*}

% table 2 has been updated
\begin{table*}
 \centering
\setlength{\tabcolsep}{0.5em}
\begin{tabular}{C{0.8cm}|C{1.5cm}C{1.5cm}C{1.5cm}|D{,}{\,\pm\,}{3}C{1.3cm}C{1.3cm}|C{1.3cm}C{1.3cm}C{1.3cm}C{1.3cm}}\hline
\multicolumn{1}{c|}{} & \multicolumn{3}{c|}{2C} & \multicolumn{3}{c|}{1C} & \multicolumn{4}{c}{2C} \\\hline
EEL & $m_V$ (mag) & $m_I$ (mag) & $m_{K'}$ (mag) & \multicolumn{1}{c}{$R_e$ (kpc)} & $n$ & $q$ & $R_e$ (kpc) & $n_{env}$  & $n_{bulge}$ & $B/T_I$   \\\hline
J0837 & $ 21.31 \pm 0.02 $ & $ 19.63 \pm 0.02 $ & $ 18.07 \pm 0.03 $ & 4.42 , 0.27 & $ 4.73 \pm 0.19 $ & $ 0.50 \pm 0.01 $ & -- & -- & -- &  --  \\
J0901 & $ 22.08 \pm 0.02 $ & $ 20.48 \pm 0.02 $ & $ 19.52 \pm 0.03 $ & 3.26 , 0.19  & $ 5.11 \pm 0.14 $ & $ 0.72 \pm 0.01 $ & -- & -- & -- &  --  \\
J0913 & $ 22.12 \pm 0.02 $ & $ 19.97 \pm 0.02 $ & $ 18.21 \pm 0.03 $ & 4.68 , 0.29  & $ 4.83 \pm 0.13 $ & $ 0.55 \pm 0.01 $ & $ 4.11 \pm 0.17 $  & $ 3.13 \pm 0.34 $ & $ 6.78 \pm 1.23$ & $0.72 \pm 0.05$ \\
J1125 & $ 23.41 \pm 0.02 $ & $ 21.85 \pm 0.02 $ & $ 19.83 \pm 0.03 $ & 4.32 , 0.46  & $ 6.24 \pm 0.29 $ & $ 0.71 \pm 0.01 $ & $ 1.17 \pm 0.02 $  & $ 0.92 \pm 0.06 $ & $ 3.06 \pm 0.46$ & $0.71 \pm 0.06$\\
J1144 & $ 21.19 \pm 0.02 $ & $ 19.77 \pm 0.02 $ & $ 19.01 \pm 0.03 $ & 8.54 , 0.68  & $ 6.85 \pm 0.19 $ & $ 0.83 \pm 0.02 $ & $ 9.64 \pm 0.28 $  & $ 0.94 \pm 0.07$ &$ 4.08 \pm 0.19 $ &  $0.61 \pm 0.06$\\
J1218 & $ 21.12 \pm 0.02 $ & $ 19.59 \pm 0.02 $ & $ 17.89 \pm 0.03 $ & 6.79 , 0.33  & $ 4.66 \pm 0.09 $ & $ 0.66 \pm 0.01 $ & -- & -- & -- &  --  \\
J1323 & $ 21.83 \pm 0.02 $ & $ 19.96 \pm 0.02 $ & $ 17.35 \pm 0.03 $ & 1.82 , 0.11  & $ 4.97 \pm 0.22 $ & $ 0.51 \pm 0.01 $ & -- & -- & -- &  --  \\
J1347 & $ 22.27 \pm 0.02 $ & $ 20.91 \pm 0.02 $ & $ 19.74 \pm 0.03 $ & 3.96 , 0.33  & $ 8.51 \pm 0.34 $ & $ 0.89 \pm 0.02 $ & $ 5.39 \pm 0.49 $  & $ 1.29 \pm 0.19$ & $ 8.09 \pm 0.40 $ & $0.40 \pm 0.05$\\
J1446 & $ 22.23 \pm 0.02 $ & $ 20.71 \pm 0.02 $ & $ 18.96 \pm 0.03 $ & 2.50 , 0.09  & $ 4.13 \pm 0.09 $ & $ 0.53 \pm 0.01 $ & $ 1.59 \pm 0.04 $  & $ 0.50 \pm 0.02 $ & $ 3.98 \pm 0.23$ & $0.47 \pm 0.07$\\
J1605 & $ 22.62 \pm 0.02 $ & $ 20.44 \pm 0.02 $ & $ 18.38 \pm 0.03 $ & 3.36 , 0.13  & $ 4.16 \pm 0.09 $ & $ 0.71 \pm 0.01 $ & $ 2.56 \pm 0.05 $  & $ 1.18 \pm 0.08 $ & $ 2.73 \pm 0.31$ & $0.72 \pm 0.06$\\
J1606 & $ 21.57 \pm 0.02 $ & $ 19.93 \pm 0.02 $ & $ 17.91 \pm 0.03 $ & 15.91 , 0.42 & $ 8.40 \pm 0.11 $ & $ 0.24 \pm 0.00 $ & $ 3.12 \pm 0.12 $  & $ 0.53 \pm 0.01$& $ 7.74 \pm 0.28 $ & $0.26 \pm 0.04$ \\
J1619 & $ 21.17 \pm 0.02 $ & $ 19.64 \pm 0.02 $ & $ 18.51 \pm 0.03 $ & 7.32 , 0.73  & $ 6.17 \pm 0.23 $ & $ 0.69 \pm 0.01 $ & $ 5.24 \pm 0.20 $  & $ 1.49 \pm 0.15$ & $ 5.07 \pm 0.35 $ & $0.44 \pm 0.08$\\
J2228 & $ 21.27 \pm 0.02 $ & $ 19.60 \pm 0.02 $ & $ 18.61 \pm 0.03 $ & 12.32 , 0.77 & $ 9.41 \pm 0.19 $ & $ 0.80 \pm 0.01 $ & $ 4.15 \pm 0.08 $  & $ 0.66 \pm 0.03$ & $ 4.65 \pm 0.23 $ & $0.52 \pm 0.05$\\\hline
\end{tabular}
\caption{A summary of source galaxy properties, with statistical uncertainties. Columns 2-4 give the unlensed $m_V$, $m_I$ and $m_{K'}$ apparent magnitudes, calculated for the `best' model (i.e. 1C or 2C, as given in Table 2). Columns 5 - 7 give the effective radius, S\'ersic index and axis ratio for the one-component models. Columns 8 - 11 give the corresponding properties of the two-component models (where they exist): here, the effective radius is that containing half the total (summed) light, taking into account both components. $n_{env}$ and $n_{bulge}$ are the S\'ersic indices of the envelope-like and bulge-like components and $B/T_{I}$ is the bulge-to-total ratio measured in the $I$-band.}
\label{tab:lensmodelsX}
\end{table*}

\section{Modelling Results}
\label{sec:lmr}

The results of our lens modelling are summarised in Tables~\ref{tab:lensmodelsY} (mass models) and~\ref{tab:lensmodelsX} (source models). We omit the lens J1248 because the lensing galaxy is clearly an edge-on disky galaxy and we find that the elliptical power law plus external shear mass model does not adequately describe the lensing potential. For the sources, we present the $VIK$ magnitudes, the effective radii $R_e$, and S\'ersic indices for both 1C and 2C models, and the axis ratio $q$ for the 1C model. Since we are primarily interested in the source properties in this study, we do not include the inference on the foreground galaxy light distributions here; these will be presented in a future work. We then present the images, models and signal-to-noise residuals for each EEL in the three bands in Figure~\ref{fig:eels}. 

While our focus is to create reproducible 1C models which are easy to interpret and compare with other studies, a number of systems presented peculiar features during the modelling process which required small changes to the main model, or simply offered interesting insights into the systems. These are summarised in the Appendix. For a number of these, we also created pixellated models of the source, using techniques similar to those described in \citet{Vegetti2009}, subtracting our best parametric model for the foreground galaxy and inferring the lensing mass distribution and regularisation. Where appropriate, these are also explained in the Appendix. %below.

\subsection{Accurately modelling the EELs}
\label{sec:seels}

In some cases, the reason for the failure of the 1C model is readily apparent. J1606, for instance, is dominated by a disk but also has a very prominent bulge which the single-component model simply cannot reproduce, and the same is true for J1446's disk; more generally, we point out that the one-component models tend to fail where the surface brightness profile is particularly extended or has a low-surface-brightness envelope, in which case the S\'ersic index becomes large in an attempt to describe both the bright, compact central structure and the extended brightness at larger radii. This raises an important point: the surface brightness structures of galaxies are generally much more complex than single S\'ersic profiles, and the fact that our sources are lensed and therefore imaged with excellent resolution, given their redshifts, means that we cannot get away with overly simple models here. We test the degree of complexity that seems to be required by adding third components to our models, and find that these tend to be poorly constrained and associated with very small amounts of flux. It seems, then, that double S\'ersic profiles are adequate -- and usually necessary -- to describe a typical EEL source.

An added complication in the modelling of these systems is that the surface brightness profiles of both foreground and background galaxy are unknown, and are presumably comparable in both colour and brightness; it is therefore possible that they are degenerate. We find, however, that this is generally not the case when both are modelled simultaneously, though it is possible that modelling in which the source is first masked and the foreground light modelled separately and then subtracted could be problematic due to the small Einstein radii of these systems.

On the other hand, we do find that the robustness of the inference on the light profiles relies on carrying out the modelling using image cutouts which capture a sufficient fraction of the light, and that this fraction is surprisingly large: our final cutout radius is $\sim$ five times the effective radius of the largest S\'ersic component in the foreground+background model (typically $\sim 5''$), and we find that modelling the same system on smaller cutouts leads to systematically different inference on the S\'ersic indices, with a larger number of foreground galaxies having components with $n<1$, and the source galaxies having systematically larger $n$. Both of these cases increase the amount of light at large radii, beyond the extent of the cutout, where it cannot be penalised by data. This emphasises the necessity of modelling the full region surrounding the lens system, in spite of the small Einstein radii of the EELs.

\begin{table}
 \centering
\begin{tabular}{c|cc}\hline
 EEL & \multicolumn{2}{c}{$\log(M_{\star}/M_{\odot})$} \\
      & lens & source \\\hline
J0837 & $ 11.08 \pm 0.10 $ & $ 11.67 \pm 0.04 $\\
J0901 & $ 10.88 \pm 0.04 $ & $ 11.19 \pm 0.04 $\\
J0913 & $ 10.93 \pm 0.04 $ & $ 11.30 \pm 0.08 $\\
J1125 & $ 11.49 \pm 0.04 $ & $ 11.01 \pm 0.06 $\\
J1144 & $ 11.02 \pm 0.06 $ & $ 11.57 \pm 0.05 $\\
J1218 & $ 11.02 \pm 0.07 $ & $ 11.63 \pm 0.05 $\\
J1323 & $ 10.51 \pm 0.21 $ & $ 11.21 \pm 0.06 $\\
J1347 & $ 10.78 \pm 0.15 $ & $ 11.12 \pm 0.08 $\\
J1446 & $ 10.80 \pm 0.07 $ & $ 11.11 \pm 0.09 $\\
J1605 & $ 11.00 \pm 0.07 $ & $ 11.09 \pm 0.09 $\\
J1606 & $ 11.25 \pm 0.01 $ & $ 11.48 \pm 0.06 $\\
J1619 & $ 11.00 \pm 0.08 $ & $ 11.47 \pm 0.12 $\\
J2228 & $ 10.25 \pm 0.53 $ & $ 11.26 \pm 0.05 $\\\hline
\end{tabular}
\caption{Stellar masses for the lens and source galaxies, inferred from the photometry using the BC03 SPS models and assuming a Chabrier IMF.}
\end{table}

\begin{figure*}
 \centering
\begin{minipage}{1.0\textwidth}
% \centering
\includegraphics[trim = 10 20 20 20,clip,width=0.495\textwidth]{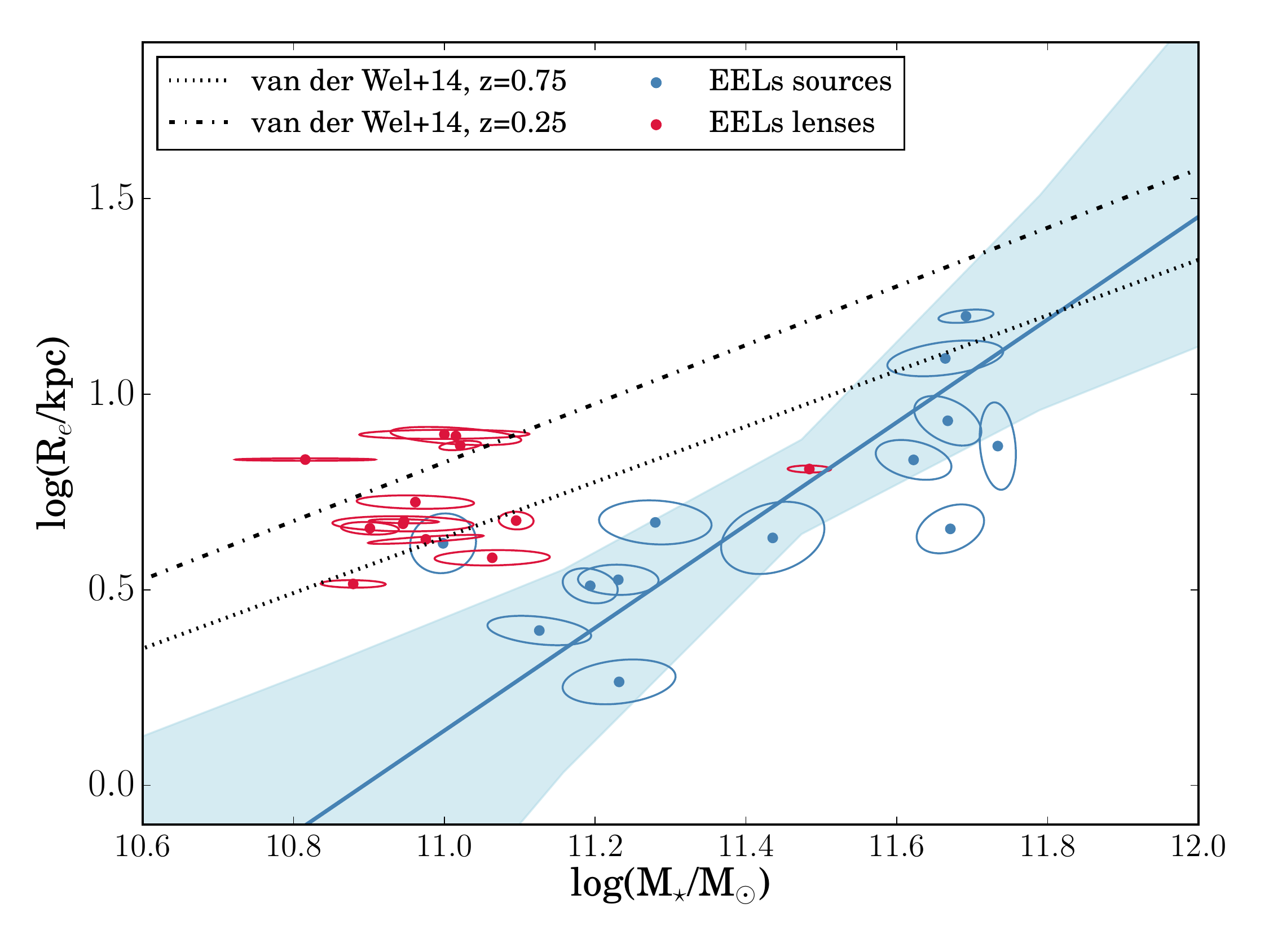}\hfill
\includegraphics[trim = 10 20 20 20,clip,width=0.495\textwidth]{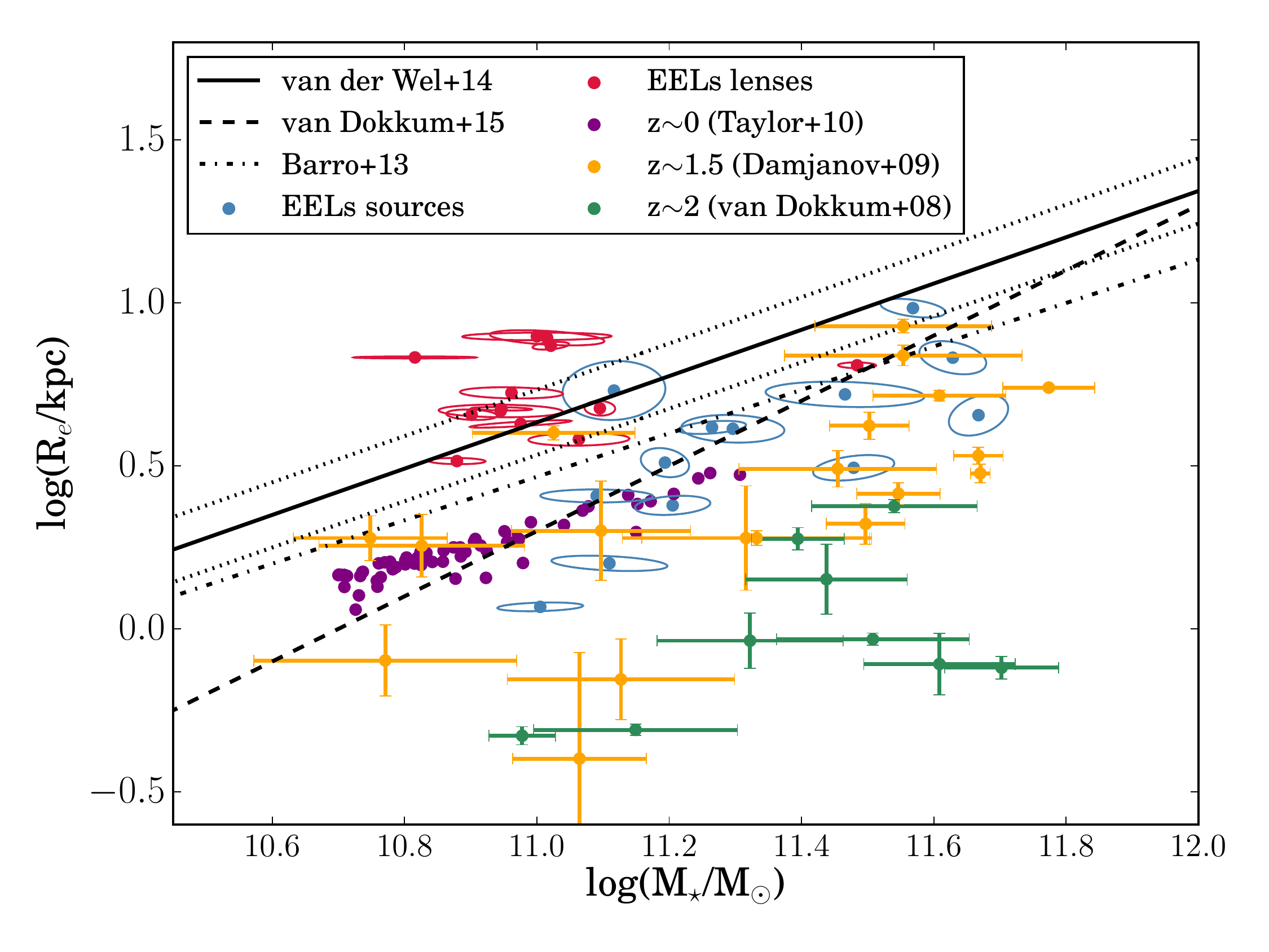}\hfill
\caption{The size-mass relation for source galaxies (blue) and lens galaxies (red). Left: 1C models, with the size-mass relations for the global ETG population from \citet{vanderWel2014} plotted for reference. The size-mass relation for the source population is well below the \citet{vanderWel2014} relation across a large part of the mass range. Right: 2C models, with the criteria for compactness used in \citet{Barro2013} and \citet{vanDokkum2015} plotted for reference in addition to the van der Wel $z = 0.75$ relation with its intrinsic scatter. Also plotted are the red nugget populations from \citet{Taylor2010}, \citet{Damjanov2009} and \citet{vanDokkum2008}, which suggest an evolution towards increasing size at lower redshifts. Our source galaxies are much more consistent with this trend within the red nugget populations, whereas the lens galaxies are consistent with the global population (though they span a very small range in stellar mass).}
\end{minipage}
\begin{minipage}{1.0\textwidth}
 \centering
\includegraphics[trim = 15 15 15 5,clip,width=0.495\textwidth]{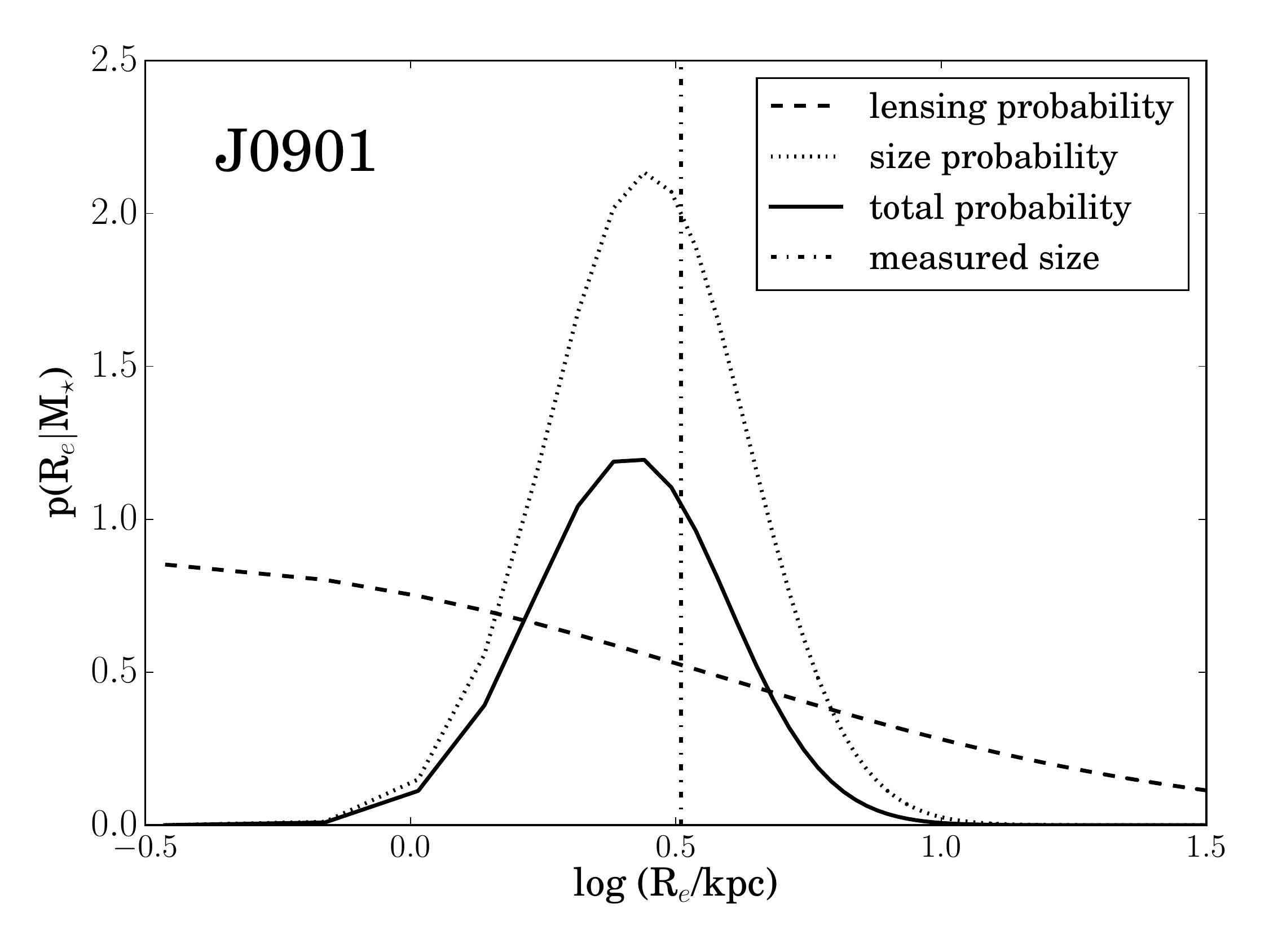}\hfill
\includegraphics[trim = 15 15 15 5,clip,width=0.495\textwidth]{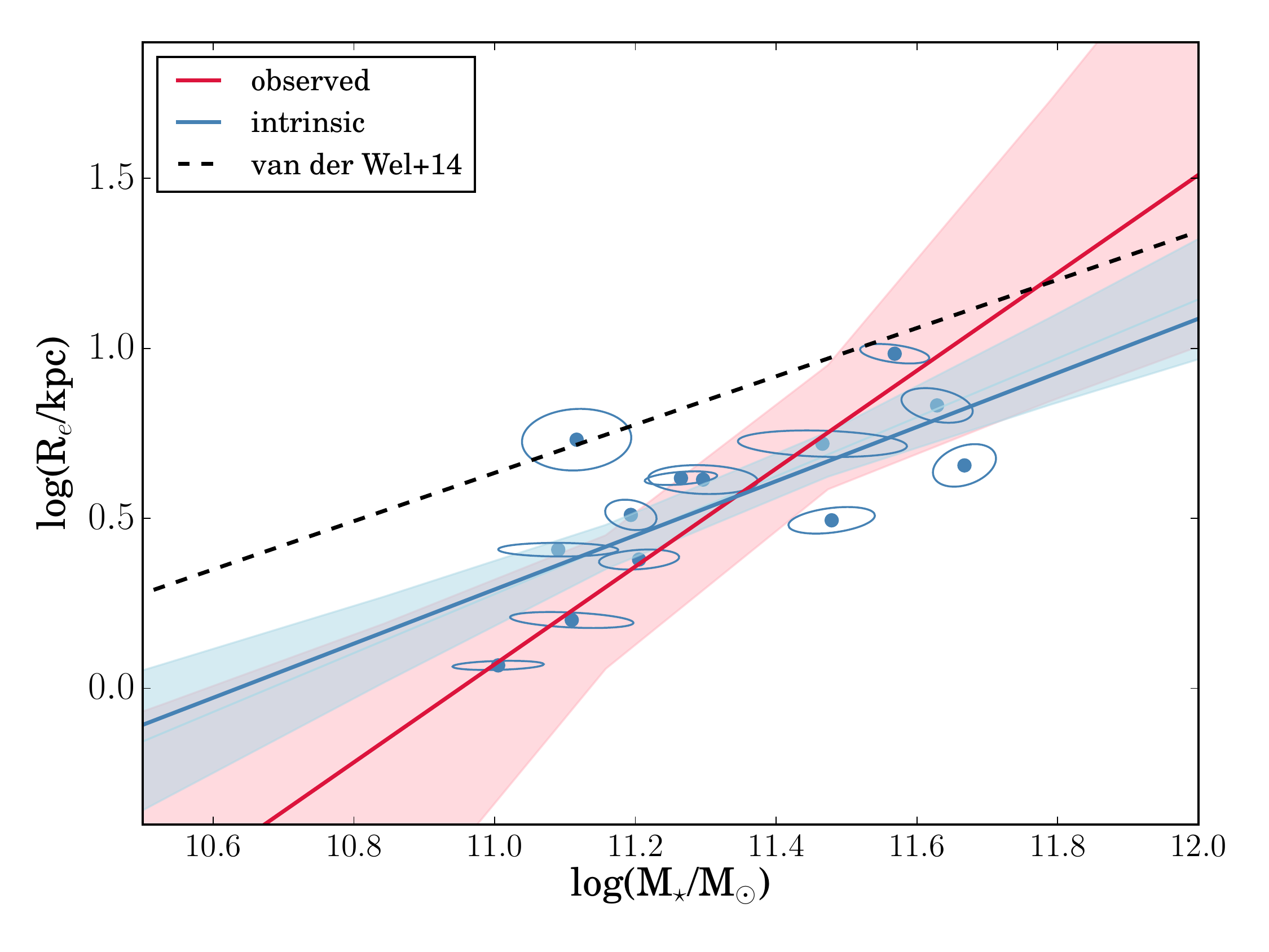}\hfill
\caption{Recovering the intrinsic size-mass distribution of compact galaxies. Left: the intrinsic size-mass relation (dotted line) at a given stellar mass is modified by the bias introduced by differential magnification (dashed line) to yield the overall probability of observing an EEL source with a particular stellar mass and effective radius (solid line). Right: The intrinsic size-mass relation (here for 2C models) is shallower than the observed relation. Relative to the $z=0.75$ \citet{vanderWel2014} relation, it is offset to smaller sizes but has a consistent slope.}

\end{minipage}
\end{figure*}

\begin{table*}
 \centering
\begin{tabular}{cccccc}\hline
 model & $\alpha_{SM}$ & $\beta_{SM}$ & $\sigma_{SM}$ & $\mu_{SM}$ & $\tau_{SM}$ \\\hline

\multicolumn{6}{c}{observed relation}\\\hline
1C & $ 0.16 _{- 0.44 }^{+ 0.27 }$ & $ 1.27 _{- 0.55 }^{+ 0.90 }$ & $ 0.11 _{- 0.08 }^{+ 0.11 }$ & $ 11.45 _{- 0.08 }^{+ 0.08 }$ & $ 0.18 _{- 0.08 }^{+ 0.10 }$ \\
2C &  $ 0.07 _{- 0.40 }^{+ 0.25 }$ & $ 1.44 _{- 0.71 }^{+ 1.15 }$ & $ 0.13 _{- 0.09 }^{+ 0.11 }$  & $ 11.33 _{- 0.08 }^{+ 0.08 }$ & $ 0.13 _{- 0.07 }^{+ 0.09 }$\\\hline %updated
\multicolumn{6}{c}{intrinsic relation}\\\hline 
1C & $ 0.36_{-0.11}^{+0.11}$ & $ 0.83_{-0.23}^{+0.22}$ & $ 0.19_{-0.04}^{+0.06}$ & $11.43_{-0.08}^{+0.08}$ & $0.28_{-0.05}^{+0.07}$ \\ 
2C & $0.28_{-0.09}^{+0.10} $ & $ 0.87_{-0.25}^{+0.24}$ & $ 0.18_{-0.04}^{+0.05}$ & $11.32_{-0.07}^{+0.07}$ & $0.24_{-0.05}^{+0.07}$ \\\hline %updated
\end{tabular}
\caption{Inference on the size-mass relation for the source galaxy population, for 1C and 2C models. The parameters correspond to those defined in Equations 3 and 4; we model the sources as following the linear relation $\log (R_e/kpc) = \beta_{SM} \log (M_{\star}/10^{11}M_{\odot}) + \alpha_{SM}$ with an intrinsic scatter $\sigma_{SM}$ in the $\log R_e$ direction, and allowing the masses to be drawn from an underlying Gaussian distribution $p(\log M_{\star}) = N(\mu_{SM}, \tau_{SM}^2)$. }
\label{tab:SMrelations}
\end{table*}

\section{Source galaxy properties}

The combination of high-resolution imaging with the magnification due to lensing means that the EELs sources can be resolved in great detail. In this Section, we present inference on their stellar masses and their size-mass relation, and point towards some characteristic features in their morphologies relative to those of the low-redshift SDSS galaxy population.

\subsection{Stellar masses}
\label{sec:sps}

As the EELs were originally identified in SDSS, each combined source+lens system also has measured $ugriz$ photometry in the SDSS database, and we can use this in addition to our $VIK$ photometry to make inference on the physical properties of both source and lens. We do not use their 2MASS photometry, as this gives little extra information alongside our NIRC2 photometry (which also has the advantage of giving magnitudes for lens and source separately, unlike the 2MASS and SDSS photometry, and thus helps to break the degeneracy between source and lens light). We also reject the SDSS $u$-band photometry, as it has very large uncertainties due to the lack of flux from ETGs at such blue wavelengths. Note that, for objects with $VIK$ photometry based on two-component models, we infer \textit{total} stellar masses using the total magnitudes, rather than assigning each component its own mass; this is because our S\'ersic profiles are only parameterisations of the light distribution and do not necessarily represent two distinct physical components.

We then infer the stellar masses of both source and lens galaxy using the composite stellar population synthesis models of \citet[BC03,][]{Bruzual2003}. Our code uses these models to compute the magnitudes, for a specified set of filters and redshift, on a grid of stellar age $T$, metallicity $Z$, dust extinction $\tau_v$ and time constant $\tau$ of an exponentially decaying star formation history, and constructs a spline interpolation model which allows magnitudes to be evaluated at arbitrary points within the grid. In this approach, we follow the methods developed by \citet{Auger2009}. We then explore the posterior probability distribution of these parameters, along with the stellar masses of the two objects, by MCMC sampling, noting that, as we are combining photometry for the separated source and lens light (from HST and Keck) with photometry for the combined system (from SDSS), the likelihood is non-linear in the logarithms of the lens and source masses $M_*$. We use uniform priors on $T$, $\tau$, $\log \tau_v$, $\log Z$ and $\log M_*$ for each object and model the source and lens photometry simultaneously, as stated previously. As discussed by \citet{Auger2009} and \citet{Newton2011}, despite large degeneracies between a number of the parameters -- such as $T$ and $Z$, and $T$ and $\tau$ -- the stellar masses are not significantly affected by these degeneracies and this makes it possible to constrain them with uncertainties of $\sim 0.05-0.1$ dex for a given IMF. We adopt a Chabrier IMF, in keeping with previous studies of the size-mass relation \citep[e.g.][]{Shen2003,vanderWel2014}, but note that the use of a Salpeter IMF -- which recent evidence suggests may be more suitable for massive ETGs \citep{Auger2010,Conroy2012} -- would increase the stellar masses by a factor of $\sim$1.7. The `best model' (i.e. 1C/2C) stellar masses for both sources and lenses are presented in Table 4.

\subsection{The observed size-mass relation}
\label{sec:smr}

We use the half-light radii inferred from the lens modelling and the stellar masses inferred from the photometry to construct the size-mass relation for both 1C and 2C models for the EELs sources. In this Section, we model the \textit{observed} relation, ignoring the selection function of the sample; we then attempt to recover the \textit{intrinsic} size-mass relation in the following Section.

We model the observed size-mass relation of the source population as a normal distribution,
\begin{equation}
 \log(R_e/\text{kpc}) = \mathcal{N}\big(\beta_{SM} \log(M_{\star}/10^{11}M_{\odot}) + \alpha_{SM}, \sigma_{SM}^2\big),
 %\log(R_e/\text{kpc}) = \beta_{SM} \log(M_{\star}/10^{11}M_{\odot}) + \alpha_{SM}
\label{eq:smr}
\end{equation}
accounting for covariance between the size and stellar mass measurements, and treating the masses as being drawn from an underlying normal distribution with mean $\mu_{SM}$ and standard deviation $\tau_{SM}$, 
\begin{equation}
 p(\log M_{\star}) = \mathcal{N}(\mu_{SM},\tau_{SM}^2).
\end{equation}
This is consistent with the fact that, as a result of the EELs selection algorithm and the galaxy mass function, we do not expect the parent distribution of stellar masses $p(\log M_{\star})$ to be flat. In this approach we follow the formalism presented by \citet{Kelly2007}. We note that while, in what follows, we model parent distributions using single normal distributions, we have verified that our inference is robust against increases in the number of normal distributions used.

The inferences for both 1C and 2C models are summarised in Table~\ref{tab:SMrelations}, and the relations are shown in Figure 2. For comparison, we also show the EELs foreground lensing galaxies, though it is clear from the figure that this population lacks the dynamic range in stellar mass to allow us to identify any meaningful trends. It is interesting to note that the sources have a larger mean mass than the lenses; we find $\mu_{SM} = 11.03$ (in units of $\log (M_{\star}/M_{\odot})$) for the lens galaxies, which is 2 times smaller than the $\mu_{SM} = 11.32$ that we calculate for the 2C models of the sources. As the cross-section for strong gravitational lensing scales approximately with lensing mass, it is an expectation that the lens galaxies will form a massive population. However, large masses for the sources are not necessarily expected, and this arises here as a result of the specific selection criteria for the EELs -- that is, detecting their spectra in the SDSS fibres requires that they be bright, with (at least a magnified) flux comparable to that of the lens galaxy. This underlines the fact that the EELs sources, as well as the lenses, constitute a massive population.

It can be seen from Figure 2 that the EELs sources are compact. We also plot the fits to the size-mass relation from \citet{vanderWel2014} -- both at $z = 0.25$ and $\ = 0.75$, which are chosen to bracket the redshifts of the EELs sources -- in the left-hand panel. Nearly all the sources lie distinctly below these lines. For comparison, the lens galaxy sample straddles the $z=0.25$ size-mass relation, as might be expected given their average redshift $\bar{z}_l = 0.35$. In the right-hand panel, we show the EELs lenses and sources alongside the red nugget populations from \citet{vanDokkum2008}, \citet{Damjanov2009} and \citet{Taylor2010}, which span redshifts between $z \sim 2$ and $z \sim 0$, in addition to the compactness criteria for classifying high-redshift nuggets used by \citet{Barro2013} and \citet{vanDokkum2015} and the global $z=0.75$ size-mass relation of \citet{vanderWel2014}, along with its intrinsic scatter. Seen in this context, the EELs source population appears to occupy a region closer to the red nuggets rather than `normal' ETGs.

We note, however, that the relations shown on this plot are meant to define some sort of boundary between `compact' and `non-compact' objects, with the former all lying below it; our EELs sources are instead scattered above and below these lines. Specifically, nine out of the thirteen systems would be classed as compact according to \citet{Barro2013}'s criterion, whereas \citet{vanDokkum2015}'s slightly stricter definition reduces this to seven -- though, due to differences in the two criteria at high and low masses, these two subsamples do not completely overlap.

Given the distinct position in size-mass space of our sources, in addition to the diversity of conflicting compactness definitions that exist, we do not think it is valuable to classify our sources in this way. Rather, we simply note that they seem to be quite massive and significantly more compact than the majority of ETGs at similar redshifts, and may be better associated with the red nugget population. For instance, they may represent red nuggets at some intermediate stage of their evolution, caught in the act of accreting matter. This is a possibility we consider in more detail in Section 6.

\subsection{The intrinsic size-mass relation}

The EELs sample is subject to a non-trivial selection function which steepens the slope of the size-mass relation that is observed. We now model this to recover the intrinsic size-mass relation.
%The observed size-mass relation of the EELs sources has a steeper slope than that of the global ETG population at similar redshifts \citep[][though it is consistent at the $2 \sigma$ level]{vanderWel2014}. This is largely a result of the non-trivial selection function that our sample has been subject to, which we now model in order to recover the intrinsic size-mass relation.

The selection function of the EELs sources is threefold. Firstly, the source must be lensed by the foreground object; this relates to the cross-section for lensing. Secondly, the inclusion of an EEL in the SDSS spectroscopic sample requires the lens+source system as a whole to fulfil the criteria of the SDSS target selection algorithm \citep{Strauss2002}, which itself is non-trivial, though the main effect is that the system is bright. Finally, the EEL must pass our spectroscopic search, which is somewhat subjective but imposes criteria such as the lensed source flux being comparable to the lens galaxy flux and the redshifts of the two objects approximately satisfying $0.1 \lesssim z \lesssim 0.7$. The combination of these different conditions leads to some selection function which modifies the intrinsic population of compact galaxies to the population of EELs sources that we observe.

Of these three contributions, the latter two are difficult to quantify and should not introduce any large bias into our measurement of the size-mass relation, although they will push us to the high-mass end of the relation. On the other hand, the first -- the lensing cross section -- introduces a selection function such that we are relatively more efficient at selecting compact galaxies at lower masses. We can understand this as follows: differential magnification introduces a bias towards smaller objects (closer to the line-of-sight of the lens), whereas, for a given size, there is no bias as a function of luminosity, and therefore stellar mass (above a limit set by the latter two criteria discussed above; note also that this is not in contradiction with the well-known magnification bias, which encodes the fact that the number density of sources increases with decreasing brightness, and not that the probability of a single object being lensed increases with decreasing brightness). The result of this is that an object of fixed luminosity becomes increasingly likely to be seen in the lensed population relative to the intrinsic population as it becomes more compact.

This effect is demonstrated by the dashed curve in the left panel of Figure 3, which shows the magnification (which we treat as a proxy for the probability of lensing) for the EEL J0901 as a function of the effective radius of the source. The shape of the curve shows that the bias is towards smaller sizes (and therefore lower-mass objects). Of course, the probability of this lensing occurring in the physical Universe also depends on the \textit{intrinsic} distribution of stellar mass and size, i.e. the intrinsic distribution of compactness, which, given the stellar mass of an object, gives the probability distribution of that object having a particular effective radius and which is what we ultimately would like to infer. In the figure, our final inference on this distribution (i.e. the intrinsic size-mass relation, see below, evaluated at the stellar mass of J0901) is shown by the dotted curve, and the corresponding probability distribution of effective radii for the EEL, given that it has been observed (i.e. the \textit{observed} size-mass relation, evaluated at the stellar mass of the EEL) is shown by the solid black curve. Thus the intrinsic size-mass distribution is modified by the bias introduced by lensing due to differential magnification.

We use this setup to infer the underlying size-mass relation, given the size-mass relation that we observe. We do this using an MCMC exploration, positing an underlying size-mass relation as in Equation~\ref{eq:smr}, and using this to calculate the probability that each EEL would be observed as a function of radius. This gives a likelihood function for the $i^{th}$ EEL
\begin{equation}
\centering
\begin{split}
 \ln L_i& = -\frac{1}{2} \Big(\frac{\log r_{e,i} - \beta_{SM} \log M_{\star,i} - \alpha_{SM}}{\sigma_i}\Big)^2  - \frac{1}{2}\ln(2 \pi \sigma_{i}^2)\\
& - \frac{1}{2}\ln \Big( \frac{\log M_{\star,i} - \mu_{SM}}{\sigma_{M,i}}\Big)^2 - \frac{1}{2}\ln(2 \pi \sigma_{\log M,i}^2)  - \ln F_i(r_{e,i})
\end{split}
\end{equation}
with dispersion for the $i^{th}$ EEL $\sigma_i^2 = \sigma_{SM}^2 + \Delta (\log r_{e,i}) ^2$ for observational uncertainty $\Delta (\log r_{e,i})$; dispersion of the underlying Gaussian distribution of stellar mass $\sigma_{M,i}^2 = \tau_{SM}^2 + \Delta (\log M_{\star,i}) ^2  $; $F_i(r_{e,i})$ is the relative magnification (i.e. the lensing probability, the dashed line in Figure 3) for the $i^{th}$ EEL at radius $r_{e,i}$, and $M_{\star,i}$ and $r_{e,i}$ are measured in units of $10^{11} M_{\odot}$ and kpc, as before. The first term here is the usual $\chi^2$ term and the second is its normalisation which must be included in the likelihood calculation as it depends on the intrinsic scatter $\sigma_{i}^2$, which is a model parameter. The third and fourth terms describe the normal distribution of the underlying parent distribution of stellar masses, and the last term accounts for the bias due to lensing.

The right-hand panel of Figure 3 shows our inference on the intrinsic size-mass relation (using the 2C models, but the 1C models yield a consistent result), and the posteriors are summarised in Table \ref{tab:SMrelations}. We find that the intrinsic slope is marginally shallower than the observed slope, and consistent with the $z = 0.75$ \citet{vanderWel2014} slope, and still offset to smaller sizes. It therefore seems that this population of compact galaxies has a size-mass relation which is systematically offset from that of the global population.

\iffalse\begin{figure*}
 \centering
\subfigure{\includegraphics[trim = 15 15 15 5,clip,width=0.48\textwidth]{J0901bias.pdf}}\hfill
\subfigure{\includegraphics[trim = 15 15 15 5,clip,width=0.48\textwidth]{intrinsic.pdf}}\hfill

  \caption{Recovering the intrinsic size-mass distribution of compact galaxies. Left: the intrinsic size-mass relation (dotted line) at a given stellar mass is modified by the bias introduced by differential magnification (dashed line) to yield the overall probability of observing an EEL source with a particular stellar mass and effective radius. Right: The intrinsic size-mass relation (here for 2C models) is shallower than the observed relation. Relative to the $z=0.75$ \citet{vanderWel2014} relation, it is offset to smaller sizes but has a consistent slope.}
\label{fig:bias}
\end{figure*}
\fi

\subsection{Morphologies}

As suggested in Section 5.2, the massive, compact nature of the EELs sources, together with their intermediate redshifts, may indicate that they are relic red nuggets, or red nuggets caught in the act of evolving. Either way, the resolving power of lensing allows us to characterise their morphologies in detail and so attempt to distinguish between different models of red nugget evolution \citep{Fan2010,Hilz2013,Ishibashi2013}. To this end, in this Section we characterise the morphologies of our EELs sources and compare them with those of the global SDSS galaxy population. Following this, we compare them with other red nuggets (the subject of Section 6.1) and predictions for red nugget growth (Section 6.2).

First, we compare the EELs sources with the global SDSS galaxy population using the bulge+disk decomposition catalogue of \citet{Simard2011}. This provides fits to a sample of roughly 1.1 million galaxies from SDSS DR7 using three different models: a pure S\'ersic model (equivalent to our 1C models), an $n_{bulge} = 4$ and exponential disk model, and an $n_{bulge}$ = free and exponential disk model (comparable, but not equivalent, to our 2C models). Specifically, we ask the question, \textit{Do the EELs sources have any distinguishing features relative to the global galaxy population?}

We find that the distributions of axis ratios and S\'ersic indices for our 1C models are both consistent with the global population. Though our sample size is small, Kolmogorov-Smirnov (KS) tests in both cases do not allow us to reject the null hypothesis that both the EELs sources and the \citet{Simard2011} galaxies are drawn from the same distribution. We note, however, that all of our sources have $n_{1C} \gtrsim 4$, which seems to indicate that all have significant bulge components -- that is, none are purely disky. These two null results are interesting in light of the finding of a high incidence of flattened and disky objects in the \citet{Stockton2014,Hsu2014} samples of low-redshift red nugget relics, and will be revisited in Section 6.1.

On the other hand, we find a much higher proportion of EELs sources needing two-component models relative to that in the \citet{Simard2011} catalogue. First, we note that nine out of thirteen ($\sim 70\%$) of our sources require two-component models in order for the data to be described down to the noise; in contrast, the \citet{Simard2011} catalogue provides a probability  $p(Ps)$ that a bulge+disk decomposition is \textit{not} needed over the pure S\'ersic model, and indicates that objects with $p(Ps) < 0.3$ may be treated as requiring a bulge+disk decomposition whilst those with $p(Ps) > 0.3$ may be considered spheroidal. We use this to classify the galaxies in their sample and find that only $\sim 20 \%$ fall into the bulge+disk category. This is particularly striking given that the \citet{Simard2011} catalogue contains spiral galaxies in addition to ETGs (they do not apply morphological cuts), whereas our EELs sources are all ETGs. This seems to indicate a significant morphological difference between the ETGs in the two samples, with our galaxies being much more likely to have a flatter, more extended component in addition to the central bulge. This is further underlined by the distribution of S\'ersic indices that we infer for our 2C model `flattened' components, for which $n_{env} < 1$ in all but three cases and $n_{env} < 1.5$ in all but one case. We do not require the second S\'ersic component to be flattened and it is entirely possible for objects to require two relatively spheroidal components, e.g. oriented at different angles or with particular combinations of $n$ and $R_e$ to reproduce their structures, so the fact that all our 2C models yield a flat component is further evidence that these objects tend to have disks or envelopes surrounding their central cores. This is a finding we will return to in Section 6.2 in the context of red nugget growth.

\section{Discussion}

%In this Section, we attempt to interpret our findings in terms of ETG evolution.

\subsection{Are the EELs sources red nugget relics?}

Previous studies of red nugget morphologies have been carried out at high redshift by \citet{vanDokkum2008}, and of intermediate-redshift red nugget relics by \citet{Auger2011}, \citet{Stockton2010,Stockton2014} and \citet{Hsu2014}. One general finding of the lower-redshift work was that large proportions of their samples required two-component S\'ersic models to describe the data satisfactorily, and that these two-component models generally implied disky morphologies. In this respect, our results are in accord: we also find nine out of the thirteen EELs to require two-component models. However, many of the two-component models of \citet{Stockton2010,Stockton2014} and \citet{Hsu2014} differ strikingly from ours in that, for nearly all their objects, the S\'ersic indices of \textit{both} components are consistently low -- with, for instance, five out of the seven systems in \citet{Stockton2014} having both components with $n<1.6$. While the \citet{Hsu2014} sample finds more of a range of morphologies -- possibly due to their larger sample size -- they also classify twelve out of their twenty-two (55 \%) objects as disk-like, with only two of these twelve exhibiting convincing bulges. The S\'ersic models of \citet{vanDokkum2008} are also in line with this, with their nine objects having generally low S\'ersic indices ranging between $0.5 < n < 4.5$ (though they do point out the uncertainty inherent in measuring galaxy structure at those redshifts). This is extremely different from what we report in Section 5.4, and suggests a paradigm in which ETGs are originally disky and become more spheroidal over time; though we have a large number of galaxies with some kind of outer envelope or disk, these are all accompanied by bulge-like components with $n > 4$, lending themselves very naturally to the interpretation that originally spheroidal galaxies, assembled at high redshift, have grown by accreting matter at large radii. 

This difference is surprising, and may suggest that the EELs sources are not drawn from the same population as these other objects. As those studies were particularly focussed on high-redshift nuggets, or relics of high-redshift nuggets, it may be the case that the EELs sources represent the more evolved counterparts of theirs. On the other hand, the difference may be the effect of our different search methods and selection criteria. In particular, the intermediate-redshift studies used IR photometry in addition to SDSS data in order to identify compact candidates, whereas we extract our compact galaxies from SDSS using strong lensing. Alternatively, it may be the case that the models in these previous studies were subject to larger uncertainties in their structural parameters than thought or that they were systematically underestimated. Indeed, \citet{vanDokkum2008} do note the difficulty in determining the morphologies of such small, distant objects; this is a problem that is still present to some extent at the redshifts probed by \citet{Stockton2014} and \citet{Hsu2014}, but which is mostly mitigated in our analysis by virtue of the fact that our sources have been lensed. At this stage, it is not possible to discriminate between these possibilities and so the picture remains complex. What is clear, however, is that compact ETGs at intermediate redshifts have a range of morphologies and may be at different stages in their evolution.

\subsection{Are the EELs sources evolving red nuggets?}

It is possible that the EELs sources are not relic red nuggets, but the descendants of red nuggets, caught in the middle of their evolution. If so, we should be able to interpret their characteristics in the context of red nugget evolution.

We have shown the EELs sources to have S\'ersic indices that are generally consistent with the global distribution -- though possibly under-representing the low-$n$ tail -- when modelled using single components, while two-component models almost always have a low-$n$ component in addition to a bulge. This is at least qualitatively consistent with the simulations of \citet{Hilz2013}, which considered the growth of ellipticals via minor mergers and found this to lead to inside-out growth, with the central density remaining relatively unaffected while matter is accreted in the outer parts, such that the bulge becomes embedded in an envelope of accreted matter.

The minor-merger-driven expansion scenario of \citet{Hilz2013} also predicts that the stars added at large radii should be metal-poor. In Section 6.3, we find negative colour gradients for nearly all the EELs sources, with the outskirts being bluer than the central regions; however, without spectral information we cannot say whether these gradients are being driven by age (with younger stars at larger radii) or metallicity (with metal-poorer stars at larger radii). It is therefore difficult to interpret this finding in the context of the action of mergers. Interestingly, one other prediction of those simulations is that the central dark matter fraction should undergo strong evolution with redshift (from $\sim 40 \%$ at $z = 2$ to $\gtrsim 70 \%$ today); estimating the dark matter fractions of our EELs sources from the stellar kinematics would be a useful further test of this scenario, and is something we plan to do in a future work.

We note that there are a number of alternative explanations for red nugget growth, including the AGN-feedback-driven scenario proposed by \citet{Ishibashi2013}, which allows radiation pressure to trigger star formation at large radii, and the quasar-driven `puffing-up' scenario proposed in \citet{Fan2008}, which has the expulsion of gas from the inner regions to the outskirts responsible for size evolution in these systems. These models may also lead to the bulge+envelope morphologies that seem to characterise the EELs sources; however, they do not as of yet make any quantitative predictions that would allow a more direct comparison with our data and we therefore do not comment on them any further here.  We emphasise that even our small sample reveals a diversity of morphologies. This may indicate that we are seeing objects at various stages in their evolution, but may also be evidence for the range of evolutionary mechanisms that are at work.

\subsection{Colour gradients and inside-out growth}

A general prediction of the hierarchical formation scenario for massive galaxies is that the gradual accretion of younger, lower-metallicity stars from lower-mass satellites should lead to negative colour gradients across the galaxy, with the central parts generally containing an old but more metal-rich stellar population compared the outskirts. This has been observed in a number of low-redshift ETGs \citep[e.g.][]{Franx1989,Peletier1990,TamuraOhta2003,Kuntschner2010,Tortora2010} and also in simulations \citep{deLucia2006,Tortora2013}. If red nuggets grow significantly in size, they should represent extreme examples of inside-out evolution. The EELs sources, which may be the descendants of these systems, therefore present an ideal opportunity to test these expectations.

Recently, \citet{Tortora2016} placed the first constraints on colour gradients in compact ETGs using their catalogue of 92 systems at redshifts $z \sim 0.2 - 0.7$, enabled by their high signal-to-noise KiDS dataset, and found preliminary evidence for negative gradients, consistent with the general ETG population. Here, we can exploit the magnification of our nuggets due to lensing to further constrain the colour gradients in our sample.

To do this, we create a new set of lens models in which all three bands are modelled simultaneously. To limit the dimensionality of the inference, we fix the mass profile of the lensing galaxy using our previous models (see Section~\ref{sec:lensmodelling}), and infer the light profiles of both lens and source and the $K'$-band PSF. We assume the lens galaxy's light profile to be the same in each band but we allow the source to have a wavelength-dependent half-light radius described by Equation 2. The location, ellipticity and position angle of the source are required to be the same in all bands (though we allow for an offset between bands due to imperfect image registration), and we use a single S\'ersic component to allow a straight-forward interpretation of the wavelength dependence of the radius. 

We find that ten out of twelve of the sources that we were able to successfully model exhibit clear negative gradients, with a sample median $\alpha = -0.45$ (and standard deviation $\sigma = 0.08$); of the remaining two objects, one (J1347) has a gradient consistent with zero and the other (J1144) has a mildly positive gradient. We were not able to find a satisfactory model for J1619 (see the Appendix) and exclude it from the analysis. A range of gradients -- mostly negative, but some positive -- was also noted by \citet{Tortora2016} and taken to indicate the range of initial conditions which can enable such objects to form; the properties of our sample underline this result, though we suggest that it may also indicate the diversity of evolutionary paths that these systems can follow. 

Interestingly, one object (J1125) has an extremely large negative gradient $\alpha_R = -1.83 \pm 0.11$, indicative of extreme changes in the stellar population as a function of radius, and therefore, potentially, a very extended period of accretion. We note that our 1C model for J1125 had a high S\'ersic index in the $K'$ band ($n = 8.40 \pm 0.98$) as compared to the HST bands ($n = 6.24 \pm 0.23$), which is consistent with a picture of the bulge being especially bright in the red, with faint, extended wings, and less bright at blue wavelengths relative to the wings. The very compact bulge size in J1125's 2C model is also interesting, and it may be that we are seeing an extreme case of inside-out growth in this system. 

\begin{figure}
 \centering
\subfigure{\includegraphics[trim = 15 15 15 5,clip,width=0.48\textwidth]{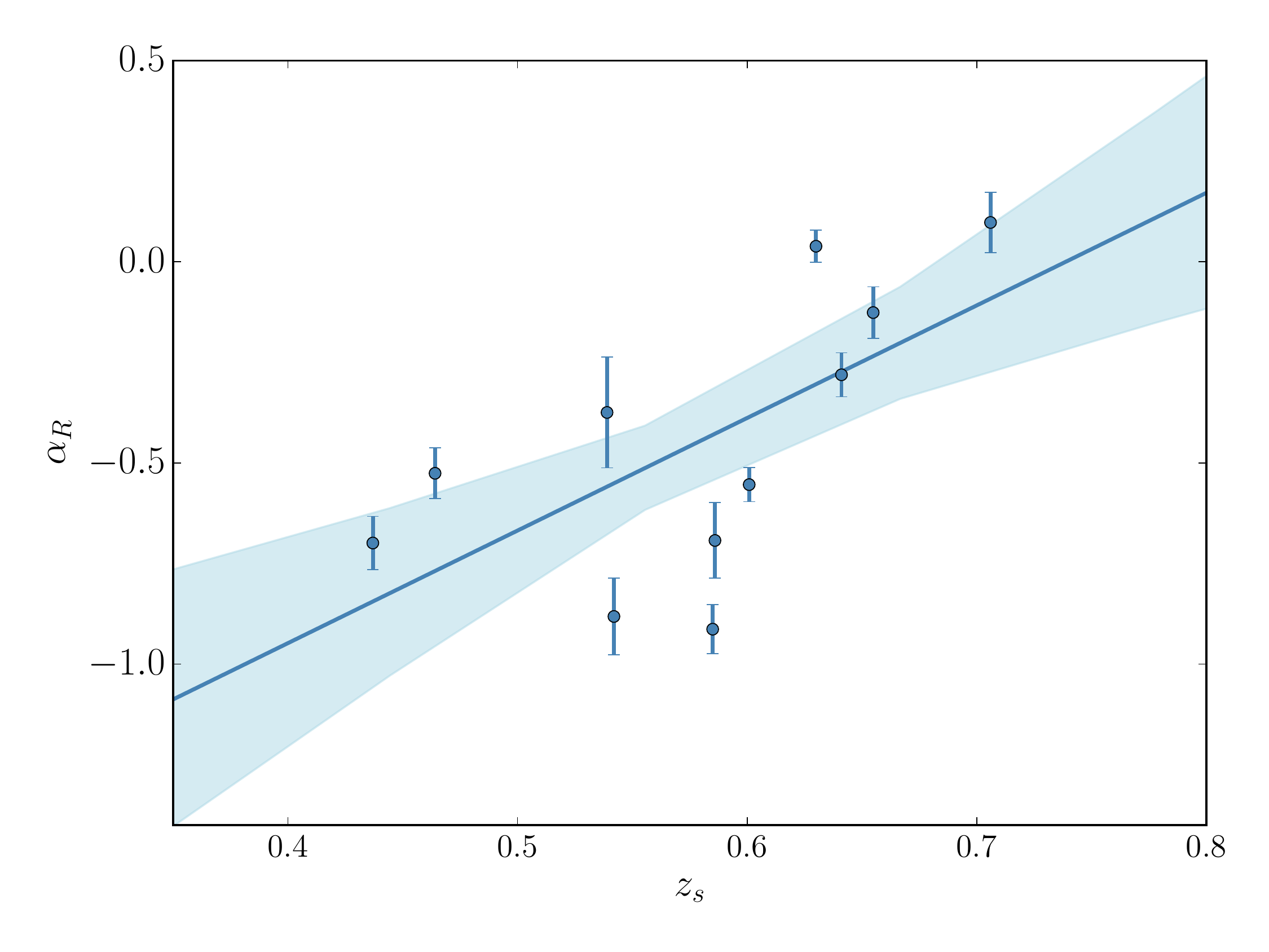}}\hfill
\subfigure{\includegraphics[trim = 15 15 15 5,clip,width=0.48\textwidth]{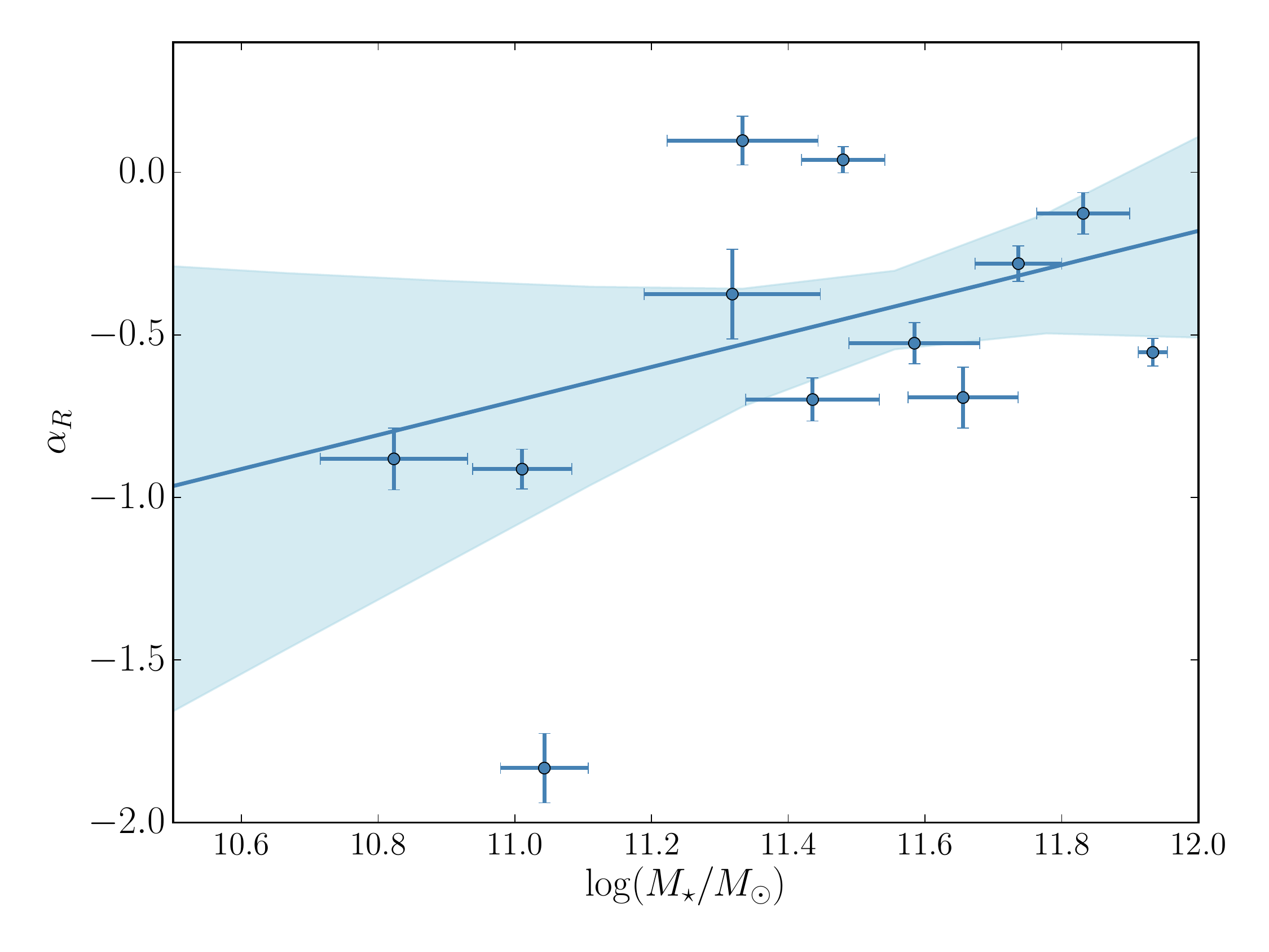}}\hfill

  \caption{Inference on relations between the radial colour gradients $\alpha_R$ and source redshift (top) and stellar mass (bottom). In both cases, we model the data as falling on a linear relation $\alpha_R = a X + b$ with some intrinsic scatter, with variable $X$ drawn from a Gaussian distribution with mean $\mu$ and standard deviation $\tau$. In both cases, we find a general trend that the radial gradients become stronger at low redshift and low mass.}
\label{fig:radgrads}
\end{figure}

Finally, we investigate the correlations of the colour gradients -- characterised by $\alpha_R$ -- with redshift $z_s$ and stellar mass $M_{\star}$, in each case modelling the correlation as $\alpha_R = a_R X + b_R$ for variable $X$ (i.e., the redshift or stellar mass) being drawn from a Gaussian distribution with mean $\mu_X$ and standard deviation $\tau_X$. As shown in the upper panel of Figure~\ref{fig:radgrads}, we find the colour gradients to be weaker ($\alpha$ less negative) at higher redshifts, and suggest that this may be because colour gradients become imprinted over time as more inside-out growth takes place. We also find that the colour gradients are weaker at higher stellar masses (lower panel); this may be the result of stellar populations in merger events mixing being more efficient at higher masses (\citealp{Kobayashi2004}, but also see \citealp{Tortora2009} for a suggestion that strong quasar feedback at high redshifts could be responsible for flattening out the colour gradients in high-mass galaxies).

\subsection{Growth in dense environments?}

A number of low-to-intermediate-redshift studies of red nuggets have suggested an important role for environment in the formation and survival of massive compact galaxies. \citet{Stringer2015} used cosmological simulations to track the evolution of a sample of compact systems and found that $94 \%$ became associated with larger structures -- either ending up embedded in clusters, or passing through such structures at an earlier phase in their lifetimes. From an observational point of view, \citet{Valentinuzzi2010a,Valentinuzzi2010b} identified a significant fraction of cluster galaxies as massive and compact, both at $0.4 < z < 1.0$ and locally at $0.04 < z < 0.07$, while \citet{Poggianti2013} found the fraction of nuggets in the field at $0.03 < z < 0.11$ to be a factor of three smaller than this (though we note that the compactness criterion used in \citet{Poggianti2013} is stricter than that in \citet{Valentinuzzi2010a,Valentinuzzi2010b}). The question arises, then, as to whether we are able to characterise the environments of the EELs source galaxies.

To that end, we investigate the SDSS galaxy population in the regions local to each source. We note here that the synthetic (i.e., as determined by the stellar population modelling of Section 5.1) $gri$ magnitudes for all thirteen EELs sources imply that they would have been detected in the SDSS $r$ and $i$ bands even if they hadn't been lensed. For each object, we query the SDSS photometric database to identify all galaxies with projected separations less than 1.5 Mpc, photometric redshifts within 0.01 of the source redshift and $0.5 < \chi^2 < 2$ for the chi-squared value of the photometric redshift; the last criterion is intended to remove objects with rogue redshifts from our count. For each source, we thus obtain an estimate of the number of galaxies which could be associated with it or become associated with it later on. We then query the database to compile catalogues of objects with similar properties to each EEL, this time using the same redshift criteria but requiring colours in the $gri$ bands to be within 0.2 magnitudes of our synthetic SDSS magnitudes for the source and imposing no cut on right ascension and declination. We call these the `twin' catalogues, and limit each one to 1000 objects. We then repeat the first step for each object in each of the twin catalogues, querying the SDSS database to estimate the number of possibly associated galaxies. This allows us to compare the distribution of associated galaxies for objects in an EEL's twin catalogue with the number of associated galaxies for the actual EEL, and so determine whether or not the EEL is residing in a particularly under- or over-dense environment with respect to other similar galaxies. We find all the EELs to be consistent with their twin catalogues, suggesting that their environments are typical of other SDSS galaxies at similar redshifts and with similar intrinsic SEDs and luminosities. This is in contrast to the suggestions of e.g.~\citet{Valentinuzzi2010a} (though see \citealp{Morishita2016} for a recent review), though we note that we cannot put strong constraints on this using photometric redshifts alone. %Nevertheless, we find no evidence that the population of EELs sources as a whole occupies over-dense environments.

\section{Summary and conclusions}

A great deal of effort has gone into explaining the evolution of compact, massive ETGs at high redshifts into the ETGs that we see in the local Universe. Proposed physical mechanisms for this growth include repeated minor merging and radiative or gas-driven AGN feedback \citep{Fan2010,Hilz2013,Ishibashi2013}, each of which makes particular predictions for the way in which these objects should evolve structurally. One of the current challenges is to identify compact objects at intermediate redshifts with which to test these predictions. We have presented a new class of ETG/ETG lenses, the EELs, and have used multiband photometry, exploiting the magnifying effect of lensing, to model the source galaxies with unprecedented resolution. These galaxies form a population of massive, compact galaxies at redshifts $z \sim 0.4 - 0.7$, and may therefore be intermediate-redshift relics of high-redshift nuggets or partly-evolved nuggets. We have carried out a survey of their structural properties so as to compare with the predictions of various models for red nugget evolution, as well as with other known or candidate low-redshift compact galaxies. Our general findings are as follows.

\begin{enumerate}
 \item The EELs sources form a massive, compact galaxy population at redshifts $z \sim 0.4 - 0.7$, lying systematically below the size-mass relation of ETGs at these redshifts.
 \item Generally, two S\'ersic components are needed to fully characterise their surface brightness distributions. This indicates complex (though smooth) morphologies and the presence of a bulge-like central component alongside a much lower-$n$ envelope-like component, both of which are compact. Indeed, two out of our thirteen objects have clear, compact envelopes. These may be the result of ongoing accretion onto the compact cores which are already in place at high redshift, in line with an inside-out formation scenario via repeated minor mergers. The diversity of structures that we observe in our small sample highlights the strong evolution that these objects undergo at intermediate redshifts.
 \item The EELs sources generally exhibit negative colour gradients, with redder centres and bluer outskirts. While we cannot disentangle the contributions from the age and metallicity of the stellar populations, we note that accretion of lower-mass galaxies with younger or lower-metallicity stars would be consistent with this trend. We also find that colour gradients are stronger at lower redshift and lower stellar mass, in line with a picture in which low-redshift galaxies have experienced more accretion and high-mass galaxies are more efficient at mixing their stellar populations.
 \item The EELs sources do not appear to occupy over-dense environments with respect to other SDSS galaxies with similar colours, luminosities and redshifts. This is contrary to suggestions that compact galaxies eventually become embedded in groups or clusters, though we cannot place strong constraints on this at present. 
\end{enumerate}

The lensing of these compact galaxies allows us to model their structures in detail and so place constraints on scenarios for their evolution. As low-redshift relics start to be discovered in increasing numbers, these constraints will be valuable in order to understand the evolving number density of these objects and the implications of this on our understanding of the local Universe. Furthermore, additional clues to their evolutionary history will be uncovered with spectroscopic observations to constrain the dynamics and stellar populations of these galaxies, and we will investigate the fundamental plane of these EELs sources in a forthcoming analysis.

\section{Acknowledgements}

We thank Tom Collett for useful discussions. LJO thanks the Science and Technology Facilities Council (STFC) for the award of a studentship. MWA also acknowledges support from the STFC in the form of an Ernest Rutherford Fellowship.  CDF acknowledges support from STScI (HST-GO-13661) and from the NSF (AST-1312329). LVEK is supported in part through an NWO-VICI career grant (project number 639.043.308). PJM acknowledges support from the U.S. Department of Energy under contract number DE-AC02-76SF00515. BJB was supported by a Marsden Fast-Start grant from the Royal Society of New Zealand, and research and study leave support from the University of Auckland.

This paper presents data based on observations made with the NASA/ESA Hubble Space Telescope, obtained at the Space Telescope Science Institute, which is operated by the Association of Universities for Research in Astronomy, Inc., under NASA contract NAS 5-26555. These observations are associated with programme GO 13661 (PI: Auger). This paper also includes data obtained at the W.M. Keck Observatory, which is operated as a scientific partnership among the California Institute of Technology, the University of California and the National Aeronautics and Space Administration. The Observatory was made possible by the generous financial support of the W.M. Keck Foundation. The authors wish to recognize and acknowledge the very significant cultural role and reverence that the summit of Mauna Kea has always had within the indigenous Hawaiian community.  We are most fortunate to have the opportunity to conduct observations from this mountain.

\appendix
\section{Individual systems} 

As explained in Section 4, some EELs presented unusual or interesting features or were not well described by 2C models. We summarise these systems here, and present pixellated source reconstructions for all thirteen systems in Figure A1.

\begin{enumerate}
 \item While the source in \textbf{J0837} appears fairly simple in the $K'$ band, the HST data reveal a clear dip across the middle of both arcs. Since this appears in both images, it is much more likely to be related to the source as opposed to any perturbations in the lensing mass \citep[e.g.][]{Koopmans2005}. We therefore assume this dip in the surface brightness to be due to a dust lane in the source, and model it using a second S\'ersic component which we require to have a negative amplitude. This significantly improves the source model, and suggests that this galaxy may have undergone a recent merger. Our pixellated reconstruction -- shown in Figure~\ref{fig:pixmodels} -- also recovers this dust lane.
 \item Neither the 1C nor the 2C model for \textbf{J1125} was able to fully account for the brightness of the lower arc of the source. This is especially apparent in the $I$ band residual image, and indicates that even a double-S\'ersic profile model may not be a good description of the source in this case. Moreover, the bulge component of the 2C model has an extremely small effective radius $R_e$ = 0.24 kpc and a high surface brightness (despite its small size, the bulge-to-total ratio in the $I$ band is still $B/T_I = 0.71$); the more `extended' component is also quite compact at $R_e = 1.49$ kpc. This suggests a bright compact source such as an AGN. Our pixellated models similarly fail to fully describe the brightest pixels in the arc; since we optimise these models for a regularisation which is constant across the image, this also seems to suggest the presence of an extremely compact central component which our regularisation may be smoothing away. It is also possible that the central component may be offset from the more extended one, either physically or due to dust obscuration. This is apparent in the slight asymmetry of the pixellated source, and may be an additional reason why our concentric parametric models cannot fully describe the data here. Indeed, when we relax this condition in our parametric model, the two source components do become offset by $\sim 1.3$kpc, though the remaining properties of both source and lens light profiles and the lensing mass profile remain consistent with those of the concentric model.
 \item  As a check on our inference on the source structure, we note that the $K'$ band image of \textbf{J1347} has been modelled previously by \citet{Auger2011}, and we compare our results for this object with the model reported in that study. As here, \citet{Auger2011} also find that a two-component fit is necessary to accurately model the surface brightness distribution, and that the inferred size of the source significantly increases when the second component is included. On the other hand, the total radius of our 2C model is $R_e = 3.96 \pm 0.33$ kpc, which is significantly larger than their $1.1$ kpc, and this difference is also seen in the inferred magnification (compare our $\mu = 5.09$ with their $\mu = 12$). This difference may be driven by differences imposed by the models or by the data, as the current analysis also includes the ACS optical data. Also, \citet{Auger2011} required the bulge component to follow a de Vaucouleurs profile with $n = 4$, whereas we left this as a free parameter and found $n = 7.86$, and this then has repercussions for the structure of the envelope component: indeed, \citet{Auger2011} finds a S\'ersic index of $n_{env} = 0.6$ which is substantially smaller than our $n_{env} = 1.44$. We also infer a power-law mass profile for the lensing galaxy with $\eta = 1.23 \pm 0.01$, which is significantly steeper than the SIE that was assumed in the earlier work. 
 \item While the prominent disk in \textbf{J1446} does not appear to be lensed and therefore seems at first glance to be associated with the lens galaxy, we find that 1C models with a single source component and two lens galaxy components (in which the second is highly flattened) are unable to provide a good description of the data. Further, close examination of the disk and the lens galaxy bulge reveals that the bulge is in fact offset from the centre of the disk by $\sim 0.1$ arcsec. When we then create 2C models for this system, we find that the second source component becomes highly flattened and the model provides a very good description of the data. We are therefore led to the somewhat surprising conclusion that the disk is in fact associated with the source galaxy. At source redshift $z_s = 0.58$, the physical size of the disk is actually rather small at $R_e = 1.69 \pm 0.02$ kpc, but because it extends beyond the Einstein radius of the lens, the tips of the disk are not lensed and retain their distinct disk-like structure. The fact that this galaxy is clearly disky is interesting in light of the various scenarios put forward for red nugget growth and the finding by e.g. \citet{Stockton2014} and \citet{Hsu2014} of a high fraction of flattened galaxies in their moderate-redshift red nugget samples (as discussed in Section 6).
 \item The source in \textbf{J1606} also exhibits a clear disk, although in this case it is almost totally lensed. Our 1C model for this system is really just a model for the bulge component and therefore provides a poor overall fit to the data; for our 2C model, we find that neither a highly flattened S\'ersic nor an exponential disk profile can provide entirely satisfactory fits to the disk component, and we therefore implement the second source component as a boxy bulge, with a highly flattened S\'ersic profile and circularised radial coordinate given by $r^c = (qx)^c + (y/q)^c$ where $c$ is a free parameter in the model, with $c<2$ indicating a diskiness and $c>2$ indicating boxiness. We find $c = 3.44 \pm 0.20$, implying that the source in this system has a strong bar-like central surface brightness distribution.
 \item While it is straightforward to find a good model for the $V$ band image of \textbf{J1619} -- where the signal-to-noise ratio is lowest -- models which describe both the $V$ and the $I$ bands tend to leave unsatisfactory residuals in both filters, with an undersubtracted ring of flux at the Einstein radius and a slightly oversubtracted bulge component. Our pixellated source reconstruction indicates a significant asymmetry in the source which may explain this as a limitation of our S\'ersic models. On the other hand, the pixellated model also has poor residuals, which suggests that the mass model may be at fault. For instance, there may also be a faint or dark perturber along the line of sight which our model does not include.

%\end{document}

\end{enumerate}

\nopagebreak

\begin{figure*}
 \centering
\subfigure{\includegraphics[trim = 20 20 20 20,clip,width=\textwidth]{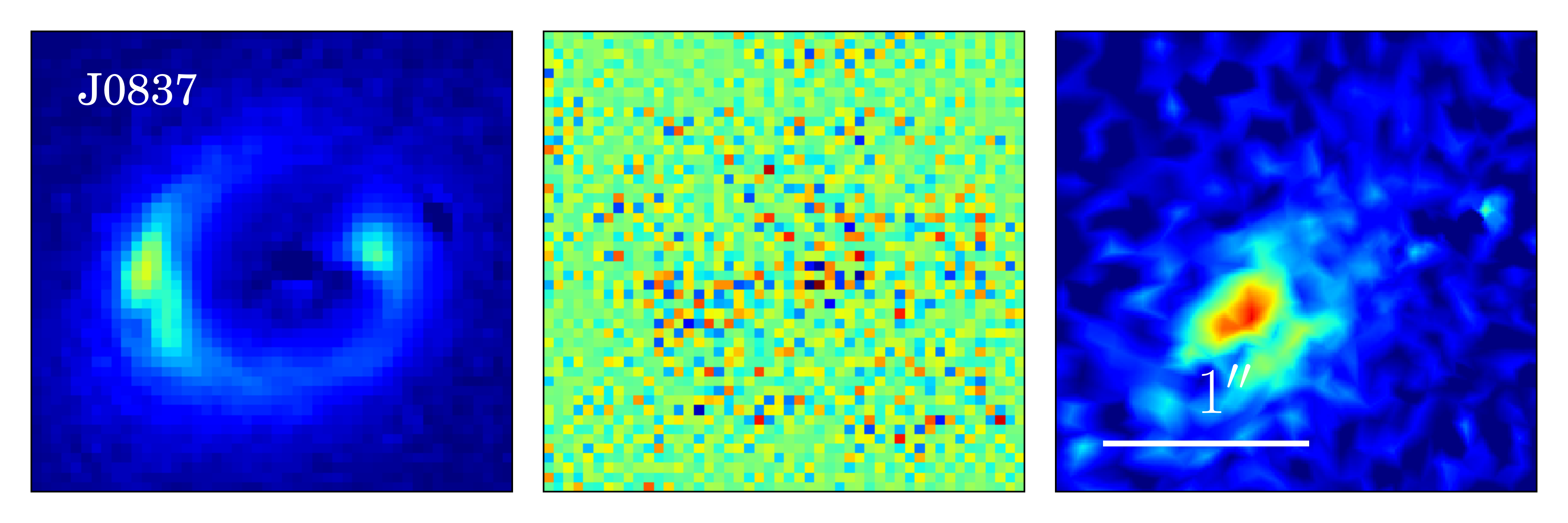}}\hfill
\subfigure{\includegraphics[trim = 20 20 20 20,clip,width=\textwidth]{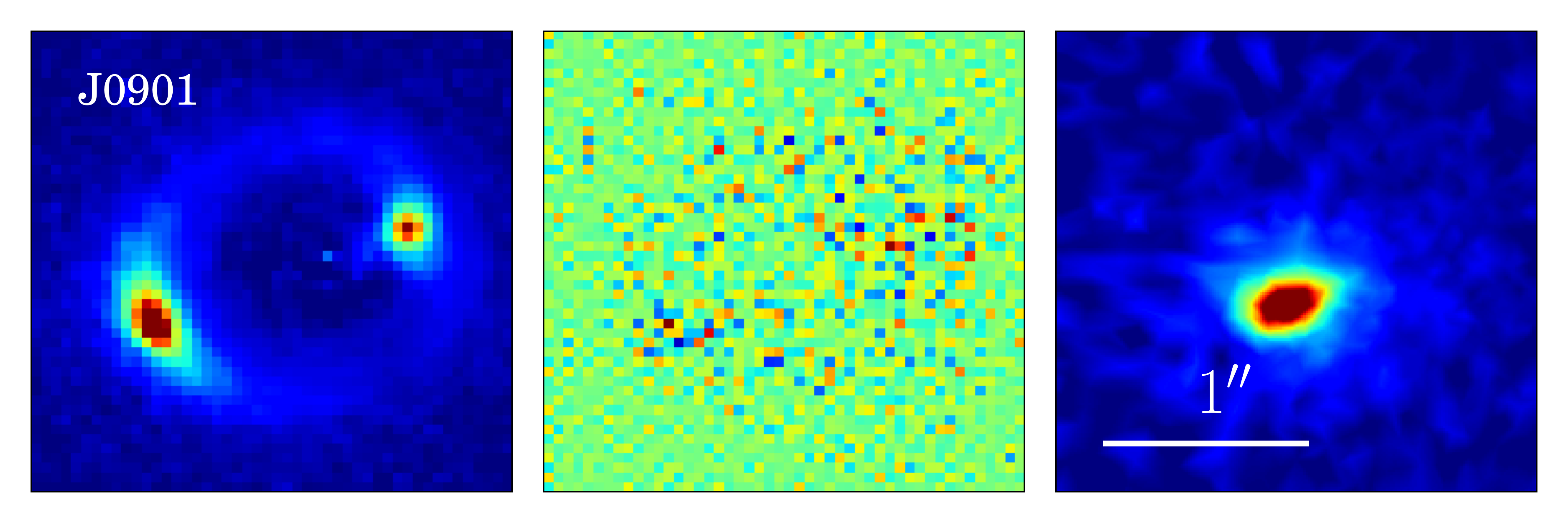}}\hfill
\subfigure{\includegraphics[trim = 20 20 20 20,clip,width=\textwidth]{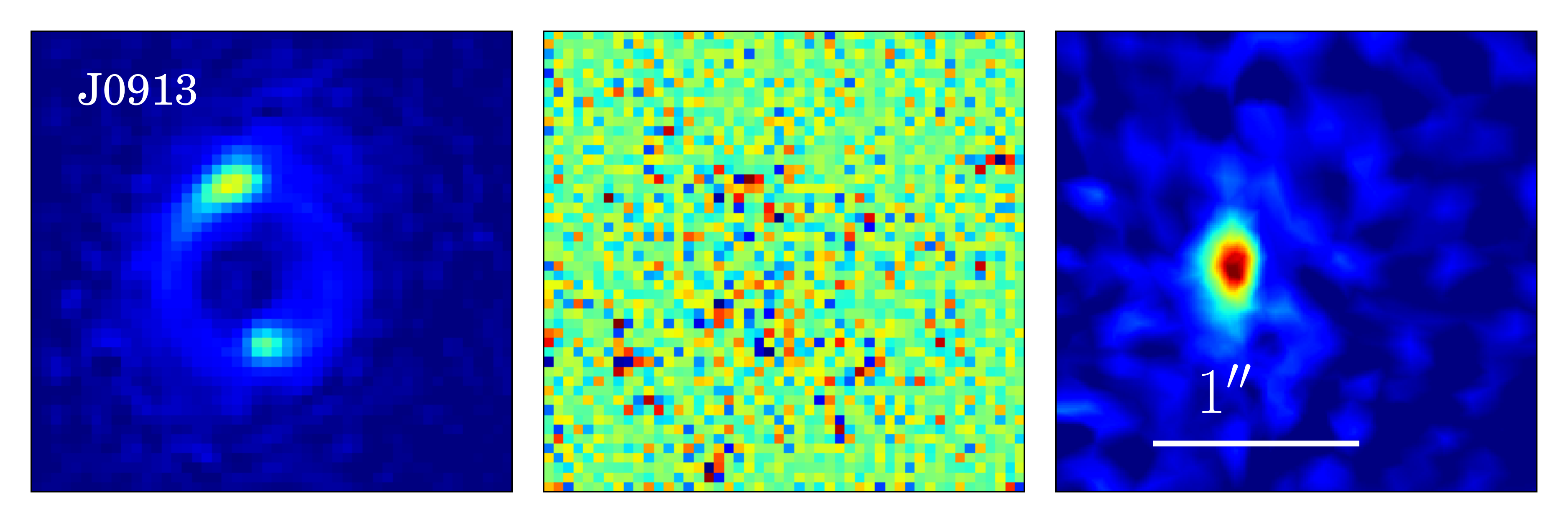}}\hfill
\caption{Pixellated reconstructions for the thirteen EELs analysed here. From left to right, we show the $V$-band image, the signal-to-noise residuals and the reconstructed source. Note that these are not fitted models, but reconstructions of the source based on the lens models inferred using parametric source models. These reconstructions generally confirm that the sources are smooth, though they also recover the dust lane feature in J0837 and the disk features in J1446 and J1606.}
\label{fig:pixmodels}
\end{figure*}

\begin{figure*}
 \centering
\subfigure{\includegraphics[trim = 20 20 20 20,clip,width=\textwidth]{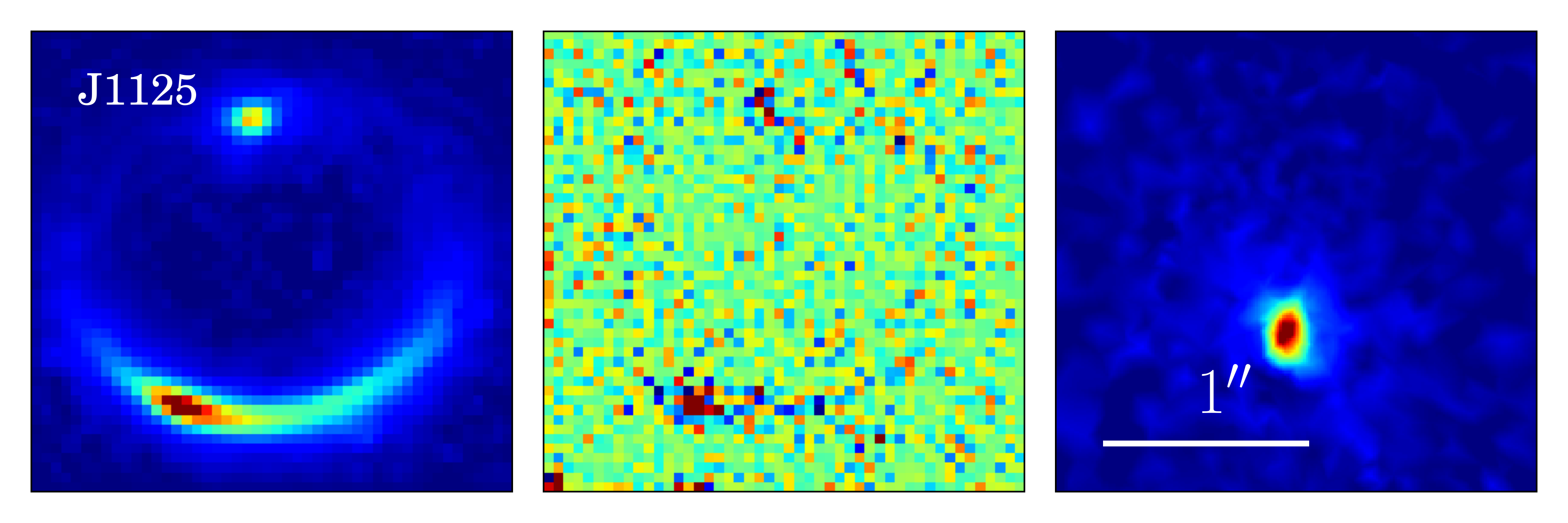}}\hfill
\subfigure{\includegraphics[trim = 20 20 20 20,clip,width=\textwidth]{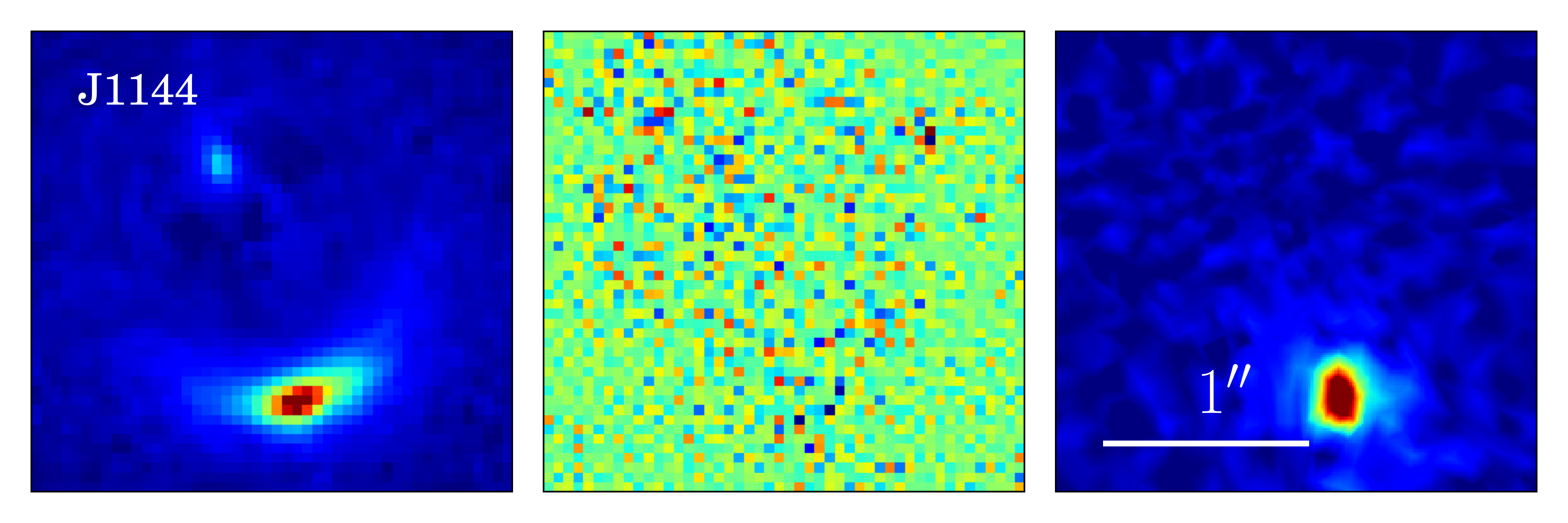}}\hfill
\subfigure{\includegraphics[trim = 20 20 20 20,clip,width=\textwidth]{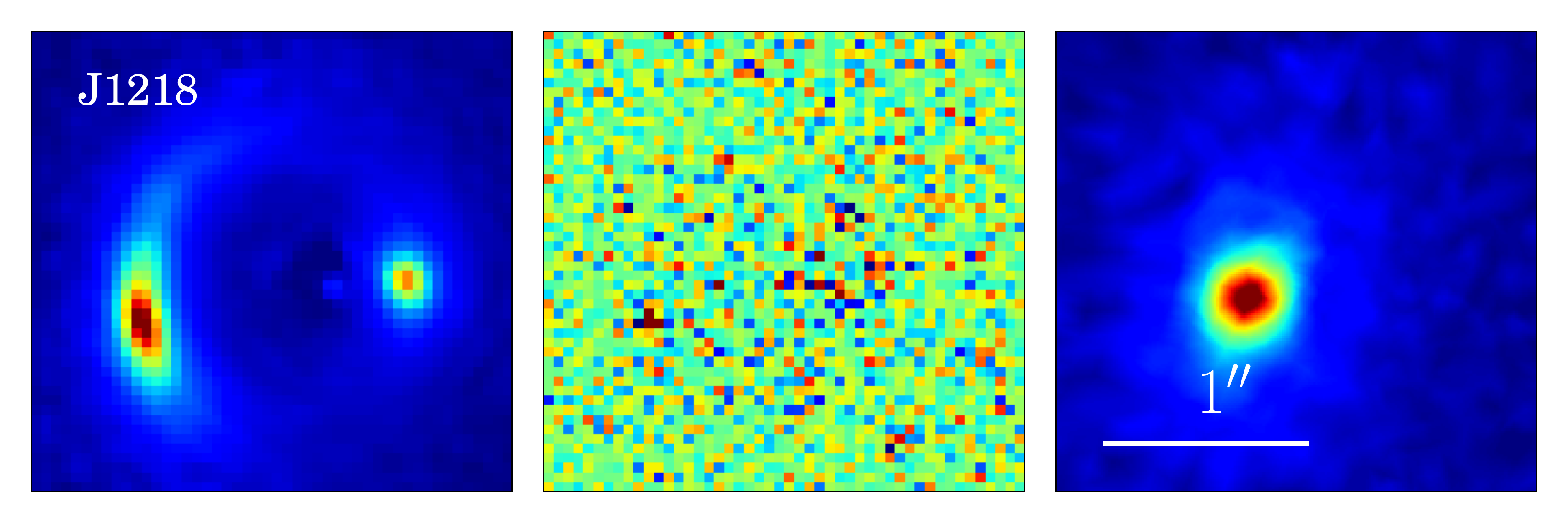}}\hfill
\subfigure{\includegraphics[trim = 20 20 20 20,clip,width=\textwidth]{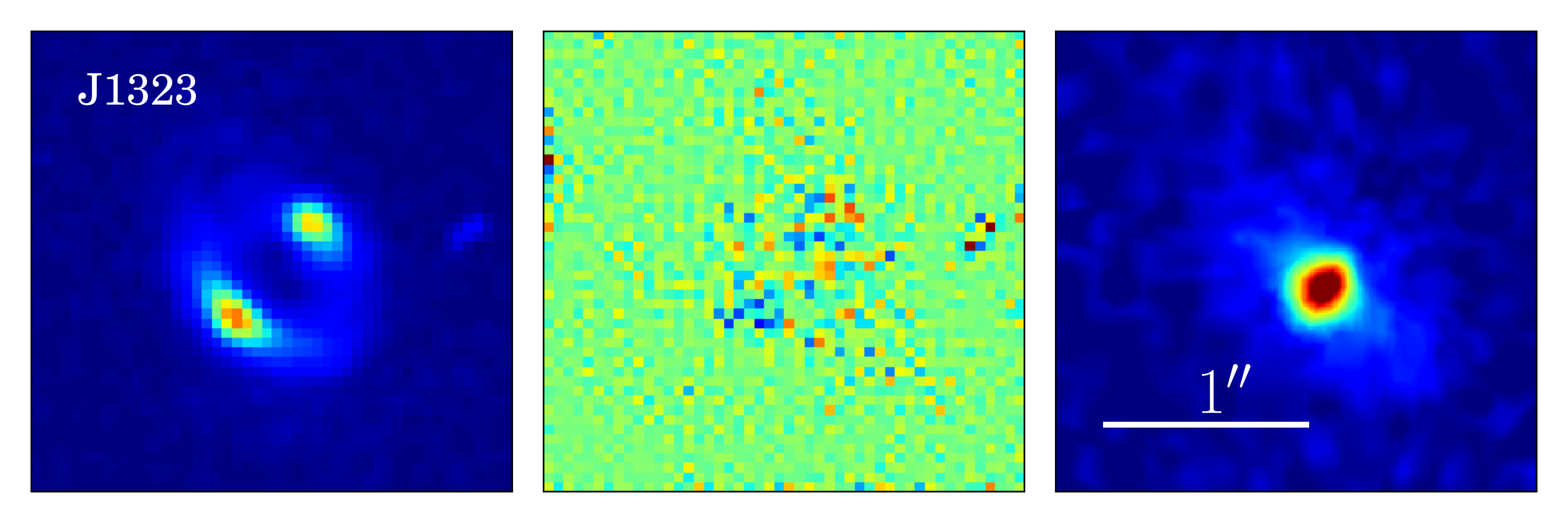}}\hfill
\contcaption{}
\end{figure*}

\begin{figure*}
 \centering
\subfigure{\includegraphics[trim = 20 20 20 20,clip,width=\textwidth]{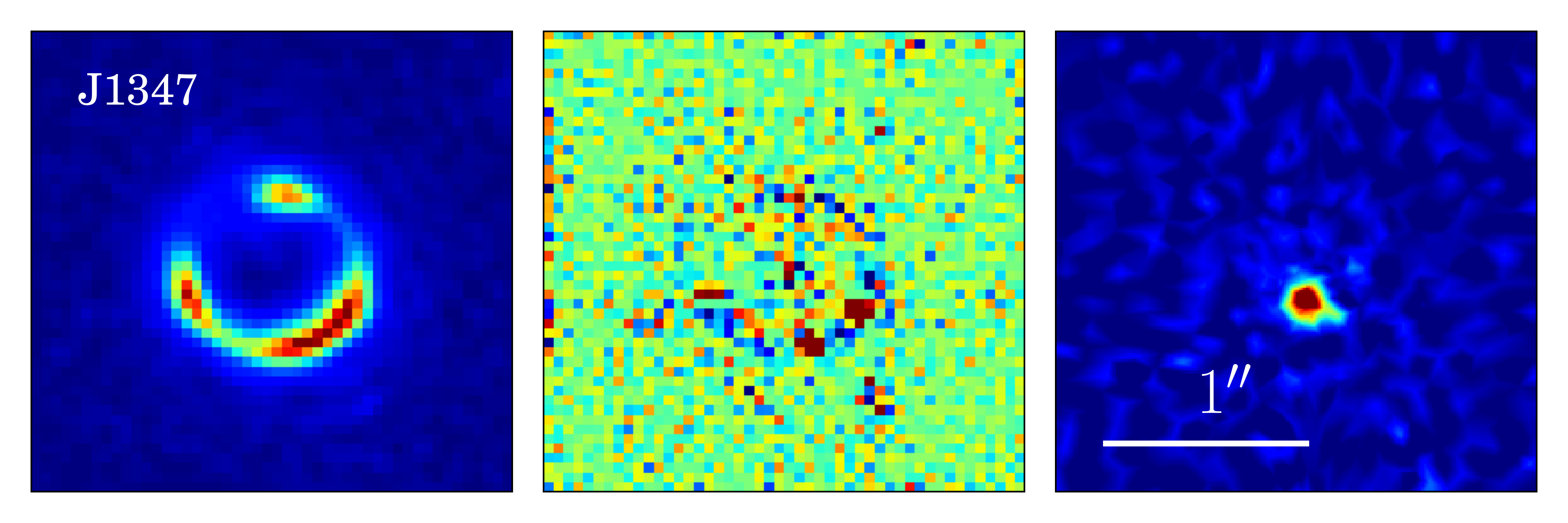}}\hfill
\subfigure{\includegraphics[trim = 20 20 20 20,clip,width=\textwidth]{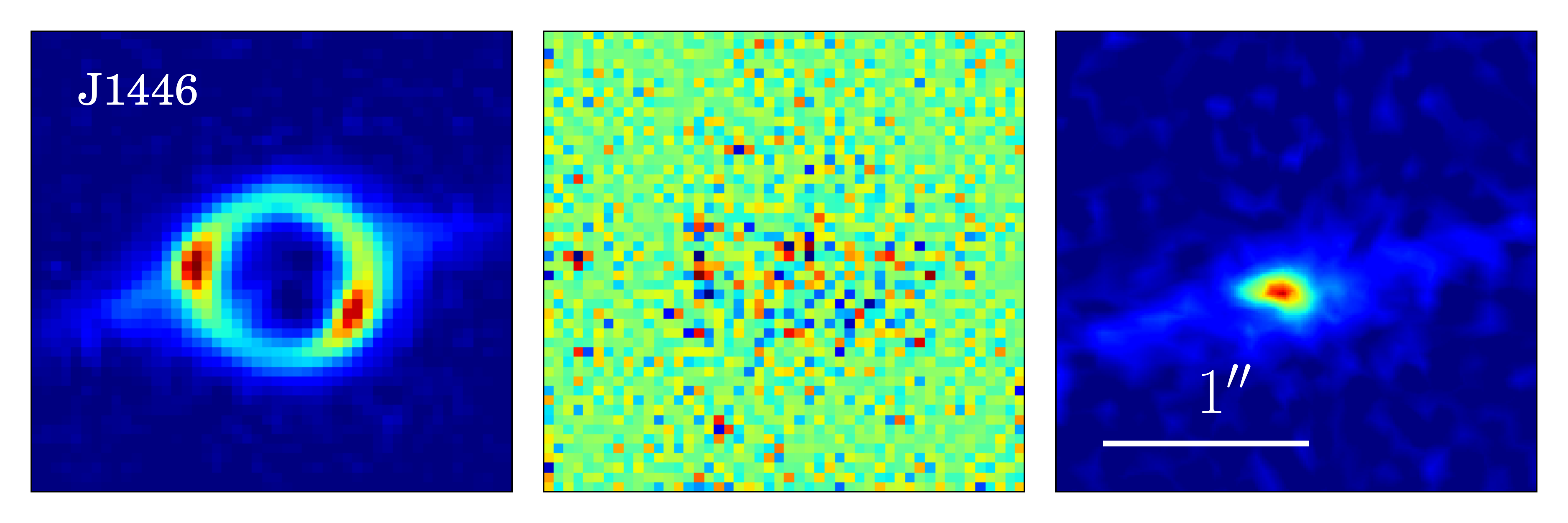}}\hfill
\subfigure{\includegraphics[trim = 20 20 20 20,clip,width=\textwidth]{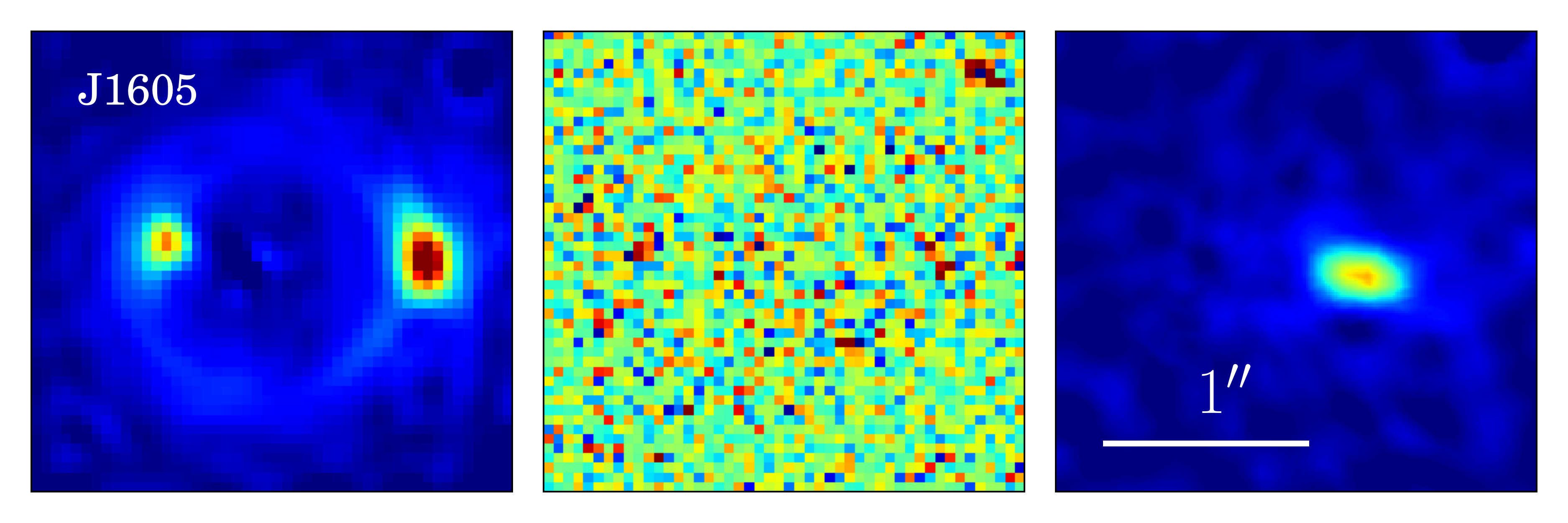}}\hfill
\subfigure{\includegraphics[trim = 20 20 20 20,clip,width=\textwidth]{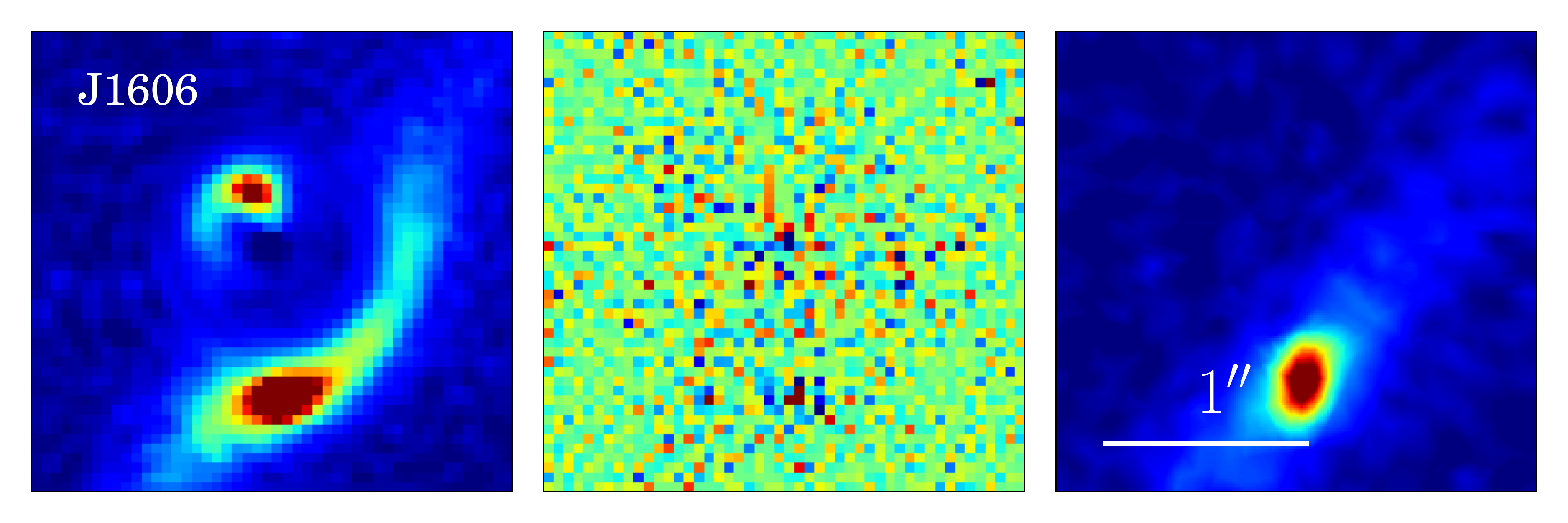}}\hfill
\contcaption{}

\end{figure*}

\begin{figure*}
 \centering
\subfigure{\includegraphics[trim = 20 20 20 20,clip,width=\textwidth]{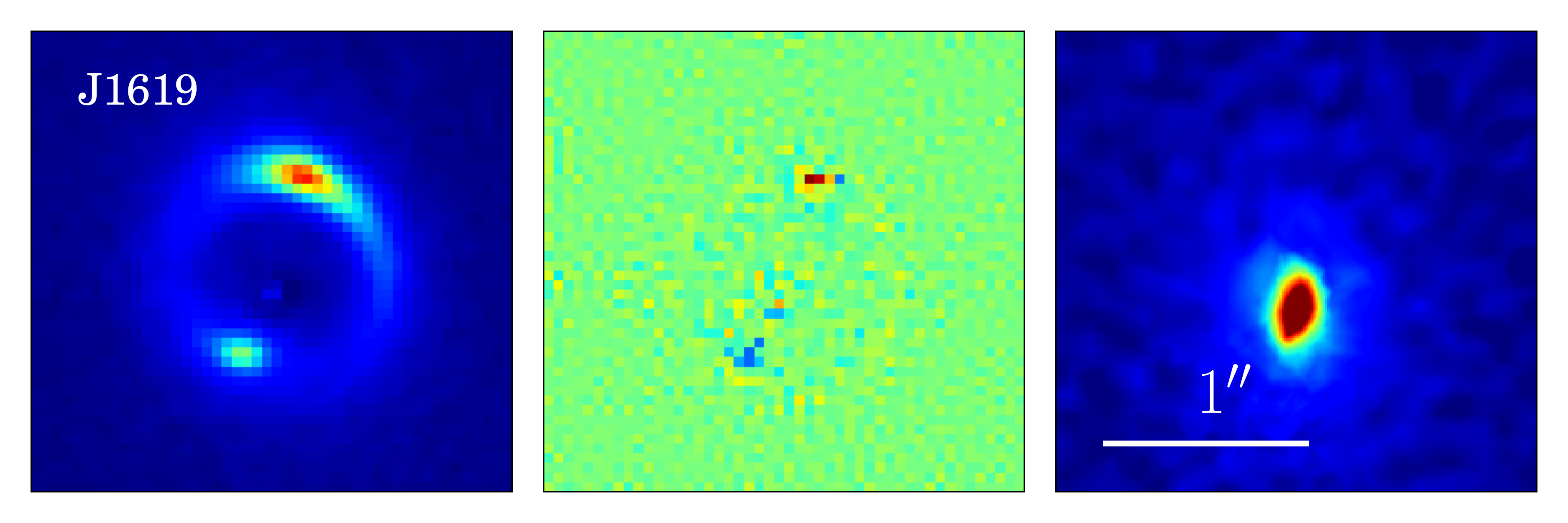}}\hfill
\subfigure{\includegraphics[trim = 20 20 20 20,clip,width=\textwidth]{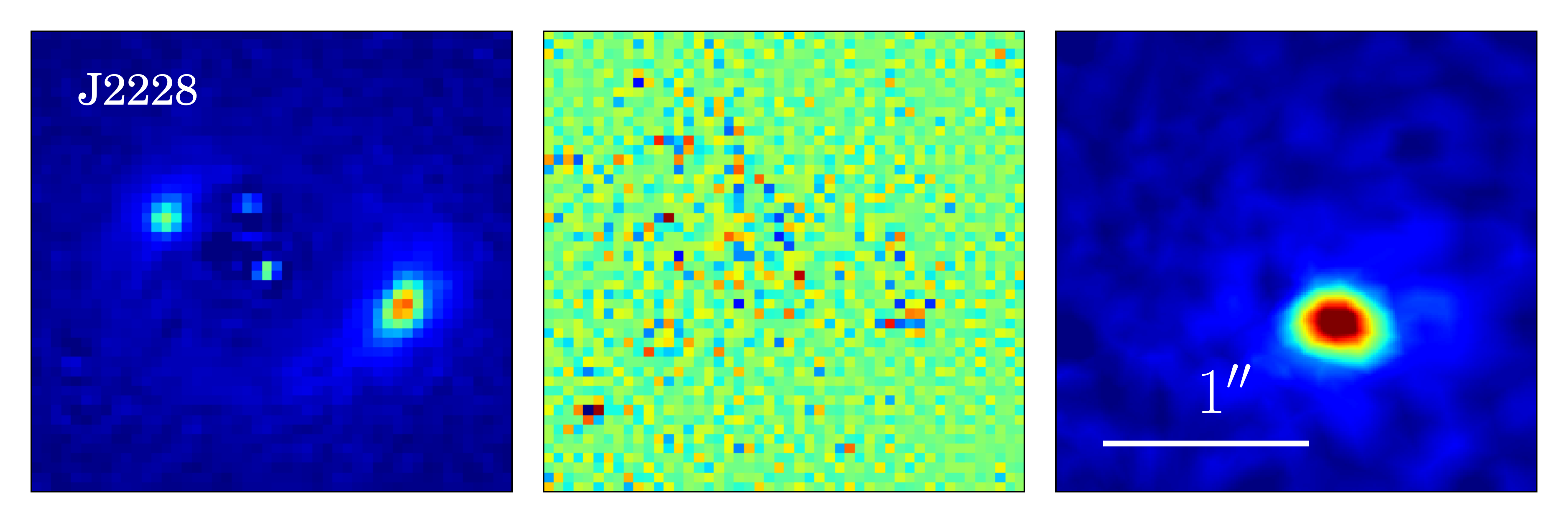}}\hfill
\contcaption{}
\end{figure*}

\end{document}